\pdfoutput=1
\documentclass{article}
\usepackage{colm2024_conference}

\makeatletter
\fancyhfoffset[L]{0pt}
\fancyhfoffset[R]{0pt}
\makeatother

\usepackage{microtype,multirow}
\usepackage{booktabs,enumitem}
\usepackage{url}
\usepackage{amsmath,amssymb}
\usepackage{caption}
\usepackage{float}
\usepackage{placeins}
\usepackage{xcolor}
\usepackage{graphicx}
\usepackage{tikz}
\makeatletter\DeclareTextCommand{\textunderscore}{OT1}{{\fontencoding{T1}\selectfont\char95}}\makeatother
\usepackage{newunicodechar}
\newunicodechar{—}{{\fontencoding{T1}\selectfont\textemdash}}
\usepackage{hyperref}
\usepackage{fancyhdr}
\usepackage{setspace}
\makeatletter
\providecommand{\github}{}
\providecommand{\huggingface}{}
\providecommand{\modelscope}{}
\providecommand{\ghlink}{}
\providecommand{\hflink}{}
\providecommand{\mslink}{}
\makeatother

\providecommand{\dtok}[1]{\texttt{\detokenize{#1}}}

\newcommand{\firstpagelogobox}{\includegraphics[height=5mm]{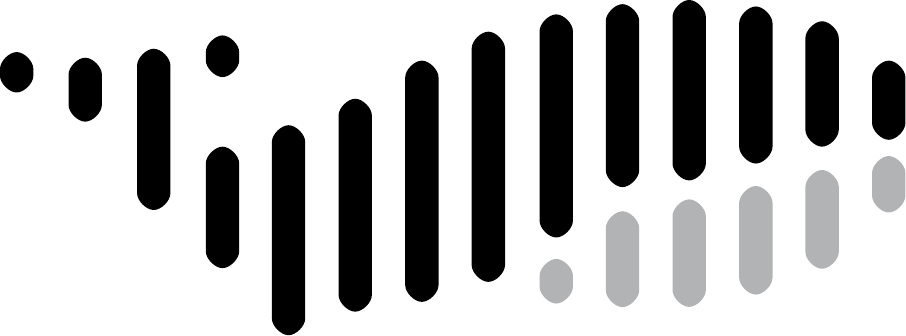}}
\definecolor{darkblue}{rgb}{0, 0, 0.5}
\hypersetup{colorlinks=true, citecolor=darkblue, linkcolor=darkblue, urlcolor=darkblue}

\title{Fish Audio Drama 3 Technical Report}
\date{}

\author{\centering Fish Audio Team}
\makeatletter
\def\@author{\vspace{-0.4em}\centering Fish Audio Team\vspace{-0.8em}}
\makeatother

\begin{document}

\pagestyle{fancy}
\fancyhf{}%
\fancyhead{}%
\renewcommand{\headrulewidth}{0.8pt}
\renewcommand{\footrulewidth}{0.8pt}
\fancyhfoffset[L]{8pt}
\fancyhfoffset[R]{0pt}
\fancyfoot[L]{}
\fancyfoot[C]{\footnotesize \strut\thepage}
\renewcommand{\footrulewidth}{0pt}

\setlength{\headheight}{30pt}
\fancypagestyle{firstpage}{%
  \fancyhf{}
  \fancyhead[L]{\raisebox{8pt}{\firstpagelogobox\hspace{6pt}\raisebox{2.8pt}{\includegraphics[height=3mm]{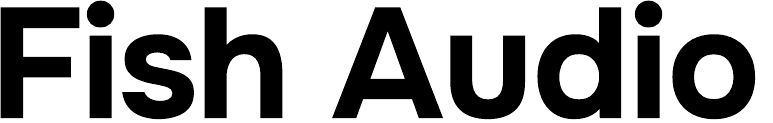}}}}
  \renewcommand{\headrulewidth}{0.6pt}
  \renewcommand{\footrulewidth}{0.6pt}
  \setlength{\headwidth}{\textwidth}
  \fancyhfoffset[L]{0pt}
  \fancyhfoffset[R]{0pt}
  \fancyfoot[L]{\footnotesize \strut\copyright\ 2026 39 AI, Inc All rights reserved.}
  \fancyfoot[C]{}
}
\setlength{\footskip}{24pt}

\vspace*{-0.3cm}
\maketitle
\thispagestyle{firstpage}
\fancyhead[L]{}
\fancyhead[C]{}
\fancyhead[R]{}
\mbox{}
\vspace{-40pt}
\begin{center}
  {\small
    \href{https://github.com/fishaudio/fish-speech}{\raisebox{-0.2ex}{\includegraphics[height=1.7ex]{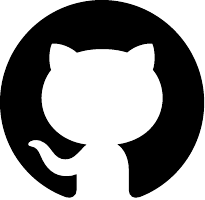}}}\quad\url{https://github.com/fishaudio/fish-speech}\par
    \href{https://huggingface.co/fishaudio/drama-3}{\raisebox{-0.2ex}{\includegraphics[height=1.7ex]{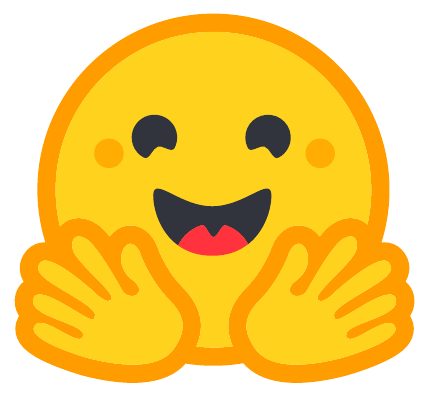}}}\quad\url{https://huggingface.co/fishaudio/drama-3}\par
    {\raisebox{-0.2ex}{\includegraphics[height=1.7ex]{logos/logo.pdf}}}\quad\url{https://fish.audio}\par
  }
\end{center}

\let\oldabstract\abstract
\let\endoldabstract\endabstract
\renewenvironment{abstract}{
  \begin{center}
    \begin{minipage}{0.88\textwidth}
    \begin{oldabstract}}{
      \end{oldabstract}
    \end{minipage}
\end{center}}

\begin{abstract}
  \vspace{10pt}

  We introduce Fish Audio Drama~3, an open-source speech generation model that turns text into directed performances, built on a 35B-parameter mixture-of-experts backbone that activates 3B parameters per token. It clones voices from multiple references, follows natural-language direction, revises a take when the next turn gives a note, voices multi-character scenes, edits existing speech by infilling, and synthesizes long-form and streaming audio.
  Multi-turn revision, scene dubbing, editing by infilling, and streaming text input are new in Drama~3, all in one model.
  Drama~3 pairs the hybrid linear-attention backbone with a new codebook multi-token-prediction audio decoder that predicts the acoustic codebooks of each frame in dedicated layers that attend over the entire sequence.
  The model is trained with a three-stage pre-training recipe, instruction fine-tuning, and large-batch reinforcement learning whose only reward is a single learned model of human preference.
  Its inference engine decodes a single stream at 271 frames per second on one H200, against 177 for Fish~Audio~S2, although Drama~3 has more than seven times as many parameters.
  The model weights are available on \textbf{\href{https://github.com/fishaudio/fish-speech}{GitHub}} and \textbf{\href{https://huggingface.co/fishaudio/drama-3}{Hugging Face}}, and inference is available through \textbf{\href{https://github.com/sgl-project/sglang-omni}{SGLang-Omni}}.
  We highly encourage readers to visit \textbf{\url{https://fish.audio}} to direct their own performances.

\end{abstract}

\section{Introduction}\label{sec:intro}

\begin{figure}[!thbp]
  \centering
  \includegraphics[width=0.94\linewidth]{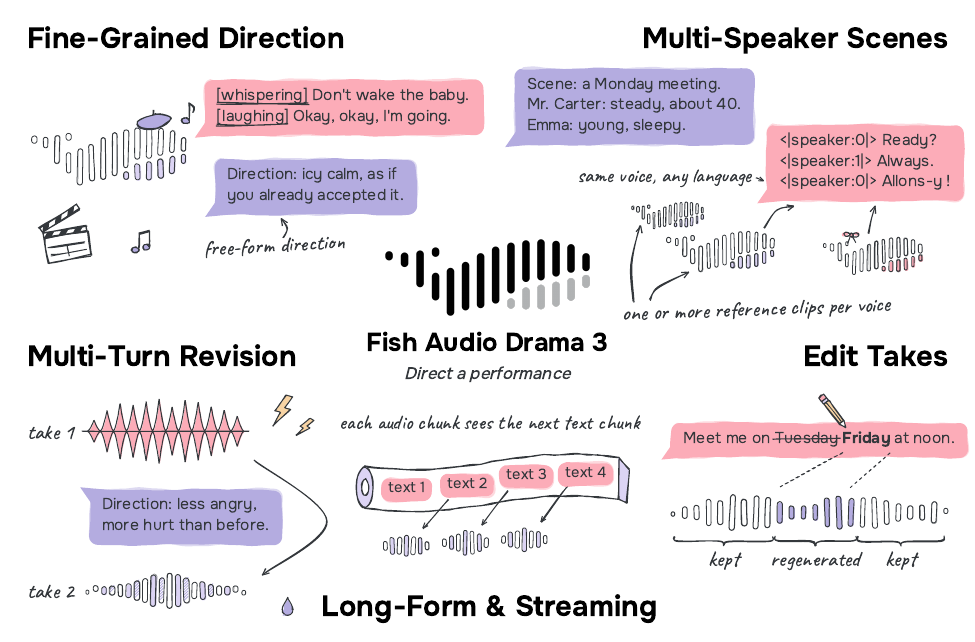}
  \caption{\textbf{Fish Audio Drama~3 turns text into directed performances.} It casts voices from references, follows natural-language direction, revises a take when the next turn gives a note, stages multi-character scenes, edits existing takes, and synthesizes long-form audio, streaming speech as the text arrives.}
  \label{fig:teaser}
\end{figure}

Large-scale text-to-speech (TTS) models now read text aloud with near-human intelligibility and speaker similarity \citep{anastassiou2024seedttsfamilyhighqualityversatile,zhang2025minimaxspeechintrinsiczeroshottexttospeech,du2025cosyvoice3inthewildspeech,hu2026qwen3ttstechnicalreport}.
The applications that drive the field today---audiobooks and audio drama, film and game dubbing, and interactive characters---ask not only for a faithful reading but for a performance.
A performance is shaped by direction, the natural-language description of how a line is to be delivered.
The delivery of a line is described in words, refined over several takes, placed within a scene among other characters, and corrected locally when a single word is wrong.

Our previous model, Fish~Audio~S2 \citep{fishaudio2026s2}, took a first step in this direction with inline natural-language instructions and native multi-speaker generation.
Four limitations remain.
First, the architecture was not built for long directed sessions, whose references and earlier takes run to thousands of audio tokens. S2's full-attention backbone attends over all of them, and its per-frame fast autoregressive module (Fast AR) \citep{liao2024fishspeechleveraginglargelanguage} sees the references only through one backbone state.
Second, control is single-shot. An instruction shapes one generation, and there is no way to give a note on a take the model has already produced and have it redone.
Third, generation is all-or-nothing. Fixing one word inside a long take requires regenerating the take, which changes everything around the word.
Fourth, post-training optimizes proxies.
Reinforcement learning (RL) for TTS commonly rewards low word error rate and high speaker similarity \citep{du2025cosyvoice3inthewildspeech,fishaudio2026s2}, yet an expressive performance legitimately departs from both. A laugh, a long pause, or a whisper changes exactly the transcript and timbre that such metrics expect.

In this report, we present Fish~Audio~Drama~3, a speech generation model built around directed performance.
Drama~3 keeps the discrete audio tokenizer of S2 and redesigns the model, the task formulation, and the post-training objective around it.
Specifically, we introduce three key technical innovations:
\begin{itemize}[leftmargin=2em, itemsep=-0.3em, topsep=-0.2em]
  \item \textbf{Hybrid Backbone with a Codebook-MTP Decoder.}
    Drama~3 builds on a hybrid linear-attention language model \citep{katharopoulos2020transformers,yang2025gated,qwen2026qwen35} with a sparse mixture-of-experts feed-forward network \citep{shazeer2017outrageously}, and replaces the per-frame Fast AR of S2 with a codebook multi-token-prediction (MTP) decoder \citep{gloeckle2024better,deepseekai2024deepseekv3}, in which each residual codebook \citep{zeghidour2022soundstream,kumar2023high} is predicted by a dedicated layer that attends causally over the entire sequence, including the fine acoustic detail of the reference audio.
  \item \textbf{Multi-Turn Revision, Scene Dubbing, Editing, and Streaming Text Input in One Model.}
    Drama~3 learns tasks that S2 did not support, such as revising a take from a note in the next turn, dubbing a whole multi-character scene, editing speech by infilling, and speaking text as it streams in.
    Each is written as a conversation over the same token stream as cloning and directing, with the loss applied only to the model's own turns (Section~\ref{sec:conv-format}, \emph{Conversation format}).
  \item \textbf{A Single Reward Model, No Rule-Based Rewards.}
    Because expressive delivery breaks the transcription and similarity proxies above, we eliminate every rule-based reward.
    Reinforcement learning optimizes a single learned reward model \citep{christiano2017deep,ouyang2022training}, a smaller speech language model of our own fine-tuned on human preference judgments, whose calibrated and bounded reward keeps reward hacking \citep{amodei2016concrete,gao2023scaling} observable.
    We train with Group Relative Policy Optimization (GRPO) \citep{shao2024deepseekmath} without a Kullback--Leibler (KL) penalty and on very large batches, which in our development was more stable than small batches with many steps.
\end{itemize}

These innovations directly enable three functional breakthroughs (Figure~\ref{fig:teaser}):
\begin{itemize}[leftmargin=2em, itemsep=-0.3em, topsep=-0.2em]
  \item \textbf{Directable Performance.} Drama~3 follows inline and scoped tags, style descriptions, and free-form director's notes. Directing is multi-turn. When a take is not what the director wanted, the next turn gives a note such as \dtok{[softer and warmer versus previous]} and the model redoes the take, with the earlier takes kept in context.
  \item \textbf{Scenes and Dialogue.} The model casts several voices from reference clips and performs multi-speaker dialogue and multi-character scenes in a single pass, with explicit control over line boundaries; a character's lines may be in different languages, in the same voice.
  \item \textbf{Editing and Long-Form Synthesis.} The model replaces a word or a sentence inside existing speech while keeping the surrounding audio, and synthesizes long-form audio line by line or in streaming mode.
\end{itemize}

Drama~3 is served by an inference engine designed for its hybrid backbone and codebook-MTP decoder (Section~\ref{sec:inference}).
Its codebook-MTP decoder runs once per frame instead of once per codebook, its experts run in the block-scaled FP8 of the RL rollouts, and the engine removes the kernel launches that CUDA graphs leave in place, optimizes kernels on the shapes of live serving, and separates prefill from decode so that new requests do not stall running streams.
On one H200, it decodes a single stream at 271 frames per second against 177 for S2, although it has more than seven times as many parameters.

We evaluate Drama~3 on public benchmarks (Section~\ref{sec:evaluation}).
For directable performance, each revision round on the Step-Audio-Edit benchmark~\citep{yan2025stepaudioeditx} raises its emotion and style accuracy.
For voice cloning, it has the lowest error rate of all compared systems in Chinese, Dutch, English, German, and Turkish on the MiniMax Multilingual Testset~\citep{zhang2025minimaxspeechintrinsiczeroshottexttospeech}.
For editing on the Ming-Freeform-Audio-Edit benchmark~\citep{yan2025minguniaudio}, it has a lower error rate than the infill API of Cartesia Sonic-3 in English and Chinese, and in Chinese a higher speaker similarity than dedicated editing models on insertion and substitution.

To accelerate research and lower the barrier to expressive speech generation, we publicly release the model weights on \textbf{\href{https://github.com/fishaudio/fish-speech}{GitHub}} and \textbf{\href{https://huggingface.co/fishaudio/drama-3}{Hugging Face}}, with inference available through \textbf{\href{https://github.com/sgl-project/sglang-omni}{SGLang-Omni}}.
To our knowledge, Drama~3 is the largest open-weight model dedicated to speech generation to date, by total parameters.
We also highly encourage readers to direct their own performances at our official site \textbf{\url{https://fish.audio/}}.

\section{Architecture}\label{sec:architecture}

\thispagestyle{fancy}
\begin{figure}[t]
  \centering
    \includegraphics[width=0.85\linewidth]{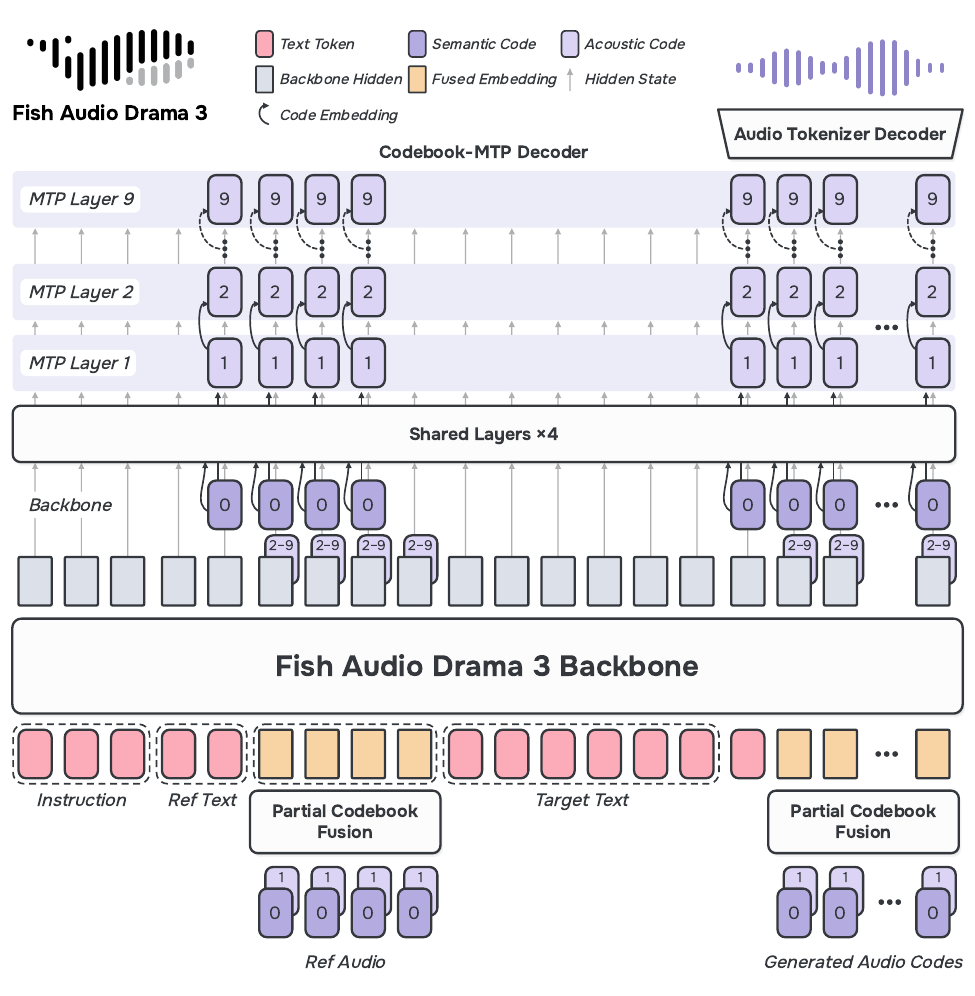}
    \caption{\textbf{Fish~Audio~Drama~3 architecture.}
    The backbone reads the system prompt, the target text, and then its own audio frames, each entered as a fused embedding of its codes 0 and 1.
    The codebook-MTP decoder runs over every position on top of the backbone's hidden state.
    Its four shared layers take that state, the predicted code 0, and the input frame's codes 2--9, which bypass the backbone, and MTP Layers 1--9 then predict the remaining codes one at a time.
    Codes are predicted wherever the next token is an audio frame, every layer attends causally along time, and the audio tokenizer decoder turns the ten codes into a waveform.}
  \label{fig:arch}
\end{figure}

Drama~3 keeps the overall design of S2, in which an autoregressive language model reads interleaved text and audio tokens and predicts the next frame's semantic code, and a second module predicts its remaining acoustic codes.
It changes three components (Figure~\ref{fig:arch}): a hybrid linear-attention backbone replaces full attention, only two of the ten codebooks are fed back into the backbone, and a time-causal codebook-MTP decoder replaces the per-frame Fast AR of S2's Dual-AR design~\citep{liao2024fishspeechleveraginglargelanguage}.
Drama~3 has no audio encoder, so reference and prompt audio enter only as audio tokens.

\subsection{Audio Tokenizer}

We reuse the S2 audio tokenizer unchanged.
Built on the Descript Audio Codec~\citep{kumar2023high}, it encodes 44.1\,kHz audio into frames of ten codebooks at about 21.5 frames per second. The first codebook is semantic, and the remaining nine are residual vector quantizers~\citep{zeghidour2022soundstream} that add progressively finer acoustic detail.
It is fully causal, so encoding and decoding both stream.
We write $q^{(k)}_t$ for the code of codebook $k \in \{0,\dots,9\}$ at frame $t$.
Each frame occupies one backbone position, so a second of speech costs about 21.5 positions regardless of the number of codebooks.

\subsection{Backbone}

The backbone is initialized from Qwen3.5-35B-A3B~\citep{qwen2026qwen35}, of which we use only the language model, a 40-layer hybrid that stacks blocks of three Gated DeltaNet layers~\citep{yang2025gated}, a gated form of linear attention~\citep{katharopoulos2020transformers, yang2024parallelizing}, and one gated full-attention layer~\citep{qiu2025gated}, a design that suits long-form speech.
The Gated DeltaNet layers carry the history in a constant-size state and only one layer in four keeps a key--value (KV) cache~\citep{pope2022efficiently}, while the full-attention layers keep exact retrieval over the whole context, such as the timbre of a reference clip or a line far back in a script.
Every layer uses a mixture-of-experts (MoE) feed-forward network~\citep{shazeer2017outrageously} that routes each token to 8 of 256 experts and adds a shared expert~\citep{dai2024deepseekmoe}, so only about 3B of the 35B backbone parameters are active per token.
We replace Qwen3.5's native MTP layer~\citep{qwen2026qwen35}, which predicts the token after next, with our own codebook-MTP decoder (Section~\ref{sec:codebook-mtp}).

\subsection{Partial Codebook Fusion}\label{sec:partial-fusion}

S2 fed all ten codebooks back into its backbone by summing their embeddings, a scheme it calls Multi-Codebook Fusion~\citep{fishaudio2026s2}.
Drama~3 feeds back only the semantic and the first acoustic codes, to limit exposure bias~\citep{bengio2015scheduled, ranzato2016sequence} from its own sampled codes while still giving the backbone a coarse view of the acoustics of both the reference audio and its own generated speech.
At the position of audio frame $t$, the backbone input is
\begin{equation}
  \mathbf{x}_t = \frac{1}{\sqrt{2}}\Bigl(\mathbf{E}^{\text{LM}}\bigl[s(q^{(0)}_t)\bigr] + \mathbf{E}^{(1)}\bigl[q^{(1)}_t\bigr]\Bigr),
  \label{eq:partial-fusion}
\end{equation}
where $\mathbf{E}^{\text{LM}}$ is the backbone's token embedding, $s(\cdot)$ maps a semantic code to one of the 4,096 semantic tokens added to the Qwen3.5 vocabulary, $\mathbf{E}^{(k)}$ is the slice of one learned codebook-embedding table reserved for codebook $k$, and $1/\sqrt{2}$ keeps the fused embedding at the scale of a single one.
Text positions use $\mathbf{E}^{\text{LM}}$ alone, and codebooks $q^{(2)}_t,\dots,q^{(9)}_t$ bypass the backbone and enter the decoder's first layer, described in Section~\ref{sec:codebook-mtp}.
In Figure~\ref{fig:arch}, the fed-back pair is the stacked 0/1 chips above Generated Audio Codes; the Partial Codebook Fusion boxes turn each pair into one fused embedding (orange), which is also how reference audio enters the token stream.

\subsection{Codebook-MTP Decoder}\label{sec:codebook-mtp}

At the position of frame $t$, the backbone predicts the next frame's semantic code $q^{(0)}_{t+1}$, and the codebook-MTP decoder predicts its nine acoustic codes $q^{(1)}_{t+1},\dots,q^{(9)}_{t+1}$ at the same position (Figure~\ref{fig:arch}, right).
The decoder stacks 13 layers of one design, with four shared layers followed by nine per-codebook layers, one for each acoustic codebook.

\textbf{Layer anatomy.}
Every decoder layer merges the hidden state from below with an injected code embedding and runs one transformer block~\citep{vaswani2017attention}:
\begin{equation}
  \begin{aligned}
    \mathbf{u}^{(\ell)}_t &= \operatorname{Merge}^{(\ell)}\bigl(\mathbf{e}^{(\ell)}_t, \mathbf{h}^{(\ell-1)}_t\bigr) = \mathbf{W}^{(\ell)}\bigl[\operatorname{RMSNorm}(\mathbf{e}^{(\ell)}_t);\,\operatorname{RMSNorm}(\mathbf{h}^{(\ell-1)}_t)\bigr], \\
    \mathbf{h}^{(\ell)}_t &= \operatorname{RMSNorm}\Bigl(\operatorname{Block}^{(\ell)}\bigl(\mathbf{u}^{(\ell)}_{1:t}\bigr)\Bigr) =: \operatorname{Layer}^{(\ell)}\bigl(\mathbf{e}^{(\ell)}_t, \mathbf{h}^{(\ell-1)}_t\bigr),
  \end{aligned}
  \label{eq:decoder-layer}
\end{equation}
where $t$ indexes positions in the whole sequence (each audio frame occupies one position, so $t$ is also its frame index), $\ell \in \{1,\dots,13\}$, $\mathbf{h}^{(0)}_t$ is the backbone's final hidden state, and $\mathbf{e}^{(\ell)}_t$ is the code embedding injected by layer $\ell$; the decoder runs over every position of the sequence, text included, and at a position that predicts no audio frame $\mathbf{e}^{(\ell)}_t$ is zero (except for the input-frame codes that the first shared layer adds, described in Equation~\ref{eq:shared-input}) and no loss is taken.
Here $\operatorname{RMSNorm}$ is root-mean-square normalization~\citep{zhang2019root}, $\mathbf{W}^{(\ell)}$ is a bias-free projection of the concatenation from $2d$ to the backbone's width $d$, and $\operatorname{Block}^{(\ell)}$ is one transformer layer with gated full attention~\citep{qiu2025gated} and a dense SwiGLU feed-forward network~\citep{shazeer2020glu} at the backbone's width, whose causal self-attention covers positions $1,\dots,t$, and $\operatorname{Layer}^{(\ell)}$ denotes the whole layer.
The MTP layers are stacked along the codebook axis instead of shifted along time, so they differ from the MTP modules that predict several future frames at once~\citep{wang2025vocalnetm2, xie2026flashtts}.
Each layer attends causally over the whole sequence, including the reference audio, so fine codes are predicted from full context rather than from one summary vector.

\textbf{Shared layers.}
At position $t$, whose input is frame $t$, the backbone predicts the semantic code $q^{(0)}_{t+1}$ of the next frame, and the decoder completes that frame at the same position.
Each of the four shared layers, $\ell = 1,\dots,4$, shown as Shared Layers $\times$4 in Figure~\ref{fig:arch}, takes the hidden state from below and injects the semantic code of the frame being completed.
The first one also receives codes 2--9 of the input frame, which the backbone never saw:
\begin{equation}
  \mathbf{e}^{(\ell)}_t = \mathbf{E}^{(\ell)}_{\text{s}}\bigl[q^{(0)}_{t+1}\bigr] + \mathbf{1}[\ell = 1]\, a_t \sum_{k=2}^{9} \mathbf{E}^{(k)}\bigl[q^{(k)}_t\bigr],
  \label{eq:shared-input}
\end{equation}
where $\mathbf{E}^{(\ell)}_{\text{s}}$ is layer $\ell$'s own table for the semantic code (the hook from code 0 into the Shared Layers box in Figure~\ref{fig:arch}), $\mathbf{1}[\cdot]$ is the indicator function, $a_t = 1$ if the input at position $t$ is an audio frame and $0$ if it is text, and $\mathbf{E}^{(k)}$ is the codebook-embedding table of Equation~\ref{eq:partial-fusion}.
The second term is the plain, unscaled sum of the embeddings of codes 2--9 (the chip stacked behind the backbone's hidden state at the foot of each audio column in Figure~\ref{fig:arch}).
A shared layer has no output head, and its only output is the hidden state of Equation~\ref{eq:decoder-layer}, which the layer above takes as its input,
\begin{equation}
  \mathbf{h}^{(\ell)}_t = \operatorname{Layer}^{(\ell)}\bigl(\mathbf{e}^{(\ell)}_t, \mathbf{h}^{(\ell-1)}_t\bigr), \qquad \ell = 1,\dots,4,
  \label{eq:shared-chain}
\end{equation}
and $\mathbf{h}^{(4)}_t$ is the input of the first codebook-MTP layer.
Together they fuse the backbone's hidden state, the semantic information of the current frame, and the acoustic detail carried over from the previous frame into a single representation.
All nine per-codebook layers then predict their codebooks from this one shared representation, with richer temporal and acoustic context than the backbone state alone.

\textbf{Per-codebook layers.}
The nine per-codebook layers (MTP Layers 1--9; $\ell = 4 + k$ for $k = 1,\dots,9$) have the same structure plus an output head, with no weights shared across codebooks.
MTP Layer $k$ takes the hidden state from below and injects the code chosen just before it, codebook $k-1$ of the frame:
\begin{equation}
  \mathbf{e}^{(4+k)}_t = \mathbf{E}^{(k)}_{\text{c}}\bigl[q^{(k-1)}_{t+1}\bigr],
  \label{eq:codebook-input}
\end{equation}
where $\mathbf{E}^{(k)}_{\text{c}}$ is the layer's own table (for $k = 1$ the code is code 0, the straight arrow past the shared box in Figure~\ref{fig:arch}).
Its hidden state follows Equation~\ref{eq:decoder-layer} as in the shared layers,
\begin{equation}
  \mathbf{h}^{(4+k)}_t = \operatorname{Layer}^{(4+k)}\bigl(\mathbf{e}^{(4+k)}_t, \mathbf{h}^{(3+k)}_t\bigr), \qquad k = 1,\dots,9,
  \label{eq:codebook-chain}
\end{equation}
starting from the shared layers' output $\mathbf{h}^{(4)}_t$; unlike a shared layer, it has an output head, which turns that state into codebook $k$ of the frame:
\begin{equation}
  q^{(k)}_{t+1} \sim \operatorname{softmax}\bigl(\mathbf{W}^{(k)}_{\text{out}}\, \mathbf{h}^{(4+k)}_t\bigr).
  \label{eq:codebook-head}
\end{equation}
Each layer thus reads the code just chosen and emits the next, up to code 9, while its attention (Equation~\ref{eq:decoder-layer}) still sees the whole sequence up to position $t$.
Reference frames receive the same injections, with their loss switched off, so every layer sees all ten codebooks of the reference audio.

\textbf{Training and inference.}
In training, every code is teacher-forced~\citep{williams1989learning} from the ground truth, so the decoder is one extra causal pass over the whole sequence, parallel over time.
At inference, each frame takes one backbone step and then 13 single-position decoder steps, and the head of each per-codebook layer is sampled before the next layer runs (Section~\ref{sec:inference}).
The sampled $q^{(0)}$ and $q^{(1)}$ of frame $t+1$ form the next backbone input (Equation~\ref{eq:partial-fusion}); its codes 2--9 go to the first shared layer as in Equation~\ref{eq:shared-input}.

\textbf{Comparison with S2.}
Codec language models handle the codebook axis in one of three ways: by delaying or flattening codebooks in the main stream~\citep{copet2023simple}; with a small depth transformer inside each frame~\citep{lee2022autoregressive, defossez2024moshi, fishaudio2026s2}; or with models that run over the whole sequence once per codebook stage, either non-autoregressively~\citep{wang2023neuralcodeclanguagemodels} or autoregressively~\citep{wang2025siren, cremer2026tontaube}.
S2's Fast AR is of the second kind, a separate four-layer transformer that loops over one frame's codebooks, conditioned on one projected backbone state~\citep{fishaudio2026s2}.
The codebook-MTP decoder is of the third kind but is one decoder rather than a chain of separate models. To our knowledge, it is the first whose per-codebook layers are full-width, attend causally along time over the backbone's own sequence, and pass the hidden state upward as DeepSeek-V3's MTP modules do~\citep{deepseekai2024deepseekv3}.
It differs from S2 in three ways.
\begin{itemize}[leftmargin=2em, itemsep=-0.3em, topsep=-0.2em]
  \item \textbf{Context.}
    It reads the fine codes of the reference audio and of all previous frames directly, whereas the Fast AR sees them only through the backbone state, so acoustic detail such as the reference's timbre reaches every codebook prediction.
  \item \textbf{Training layout.}
    It runs on the backbone's own sequence, reusing its rotary position embeddings~\citep{su2021roformer} and attention masks, so every codebook of every frame is trained in one extra causal pass, whereas the Fast AR is trained on separate per-frame codebook sequences.
  \item \textbf{Capacity.}
    It gives every codebook a full-width layer, embedding, head, and KV cache over the whole sequence of its own, whereas the Fast AR shares four layers across all codebooks and keeps only per-frame scratch state.
\end{itemize}

\section{Training}\label{sec:training}
\thispagestyle{fancy}

\begin{figure}[t]
  \centering
  \includegraphics[width=\linewidth]{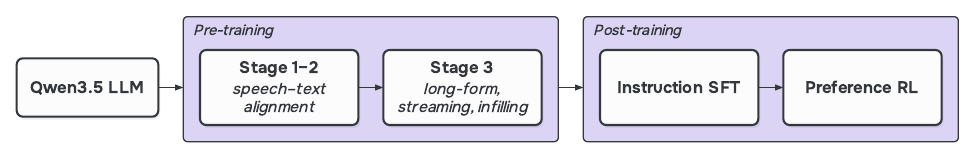}
  \caption{\textbf{Drama~3 training pipeline.} A pre-trained Qwen3.5 LLM goes through three pre-training stages, then instruction SFT and preference RL.}
  \label{fig:training}
\end{figure}

Drama~3 is initialized from a pre-trained Qwen3.5 language model~\citep{qwen2026qwen35} with a newly initialized codebook-MTP decoder (Section~\ref{sec:architecture}) and trained in three pre-training stages (Section~\ref{sec:pretrain}), one stage of supervised fine-tuning (SFT) on instructions (Section~\ref{sec:sft}), and reinforcement learning (RL) from human preference~\citep{christiano2017deep} (Section~\ref{sec:rl}), as Figure~\ref{fig:training} shows.

\textbf{Conversation format.}\phantomsection\label{sec:conv-format}
Every task is written as a conversation in the ChatML-style chat markup of the base model, in which the system turn holds the task instruction and the voice references for zero-shot cloning~\citep{wang2023neuralcodeclanguagemodels}, the user turn holds the text to speak, and the assistant turn holds the audio.
Loss covers only the assistant turns, so reference audio is context only.
Stages~1 and~2 of pre-training concatenate the references as in S2; Stage~3 either concatenates them or gives each clip a numbered block pairing its speaker-tagged transcript with its audio, at random; and from SFT on, every sequence uses these numbered blocks, the indexed reference layout (Appendices~\ref{app:pretrain-formats} and~\ref{app:prompts}).

\subsection{Pre-training}\label{sec:pretrain}

\textbf{Data.}
Pre-training data is transcribed speech, all of which passes quality control first, and text documents, which keep the backbone's language ability.
In total, pre-training processes several hundred billion tokens.
The language coverage is that of S2, about 80 languages and dialects~\citep{fishaudio2026s2}.

\textbf{Stages.}
Stages~1 and~2 teach the model to turn text into audio codes; their sequences are at most 8K tokens long and of two kinds, text documents and multi-turn TTS conversations (Figure~\ref{fig:pre-tts}).
Stage~2 continues on the same sequences with the batch doubled~\citep{smith2018dont}, which lowers gradient noise, and leaves the learning-rate decay to Stage~3.
Stage~3 extends the model to whole recordings, streaming input, and speech editing. It raises the maximum sequence length to 16K tokens and adds long-form synthesis of a whole recording from its transcript, as one segment (Figure~\ref{fig:pre-longform}) or line by line (Figure~\ref{fig:pre-lines}), streaming interleave (Figure~\ref{fig:pre-stream}), and speech infilling (Figure~\ref{fig:prompt-infill}).
Throughout, samples are packed with document masking~\citep{zhao2024analysing} into rows of 128K tokens, and the model is optimized with Muon~\citep{jordan2024muon, liu2025muon}.

\textbf{Objective.}
Let $\mathcal{T}$ be the supervised text positions, including the closing \dtok{<|im_end|>} of each audio segment and the line separators, and $\mathcal{S}$ the supervised positions whose target is a semantic code $q^{(0)}_t$; then
\begin{equation}
  \mathcal{L}_{\text{text}} = -\frac{1}{|\mathcal{T}|} \sum_{t \in \mathcal{T}} \log p_\theta(x_t \mid x_{<t}),
  \qquad
  \mathcal{L}_{\text{sem}} = -\frac{1}{|\mathcal{S}|} \sum_{t \in \mathcal{S}} \log p_\theta\bigl(q^{(0)}_t \mid x_{<t}\bigr),
  \label{eq:loss-lm}
\end{equation}
\begin{equation}
  \mathcal{L}_{\text{cb}} = \frac{1}{9} \sum_{k=1}^{9} \Bigl( -\frac{1}{|\mathcal{S}|} \sum_{t \in \mathcal{S}} \log p_\phi\bigl(q^{(k)}_t \mid \mathbf{h}_{<t}, \mathbf{q}_{<t}, q^{(<k)}_t\bigr) \Bigr),
  \qquad
  \mathcal{L} = \mathcal{L}_{\text{text}} + 0.5\, \mathcal{L}_{\text{sem}} + 0.5\, \mathcal{L}_{\text{cb}},
  \label{eq:loss-total}
\end{equation}
where $p_\theta$ is the backbone's output distribution, $p_\phi$ the decoder (Section~\ref{sec:architecture}), $\mathbf{h}_{<t}$ the backbone hidden states at earlier positions, $\mathbf{q}_{<t}$ all ten codebooks of earlier frames, and $q^{(<k)}_t$ the lower codebooks of the current frame.
Each term is a mean over its own bucket, so the text--speech balance does not drift with batch composition; the semantic and codebook terms share one per-frame mask, and $\mathcal{L}_{\text{cb}}$ back-propagates into the backbone.
In pre-training and SFT, an auxiliary load-balancing loss~\citep{lepikhin2020gshard} keeps the MoE router from sending most tokens to a few experts.

\subsection{Instruction SFT}\label{sec:sft}

A single SFT stage starts from the Stage-3 checkpoint and trains on sequences of up to 32K tokens in the indexed reference layout (Section~\ref{sec:conv-format}, \emph{Conversation format}).
It teaches the model to follow directions and to perform dialogue and scenes.
We choose the SFT recipe by the teacher-forced negative log-likelihood (NLL) of the semantic tokens of held-out real speech, each clip cloned from other clips of its voice, rather than by training loss.

\textbf{Instruction formats.}
As in S2, directions are natural language in the user text, in the forms of Appendix~\ref{app:syntax}: point tags such as \dtok{[laughs]} or \dtok{[pause]}, which insert a vocal event or a pause at that position; scoped tags such as \dtok{<whisper>...</whisper>}, which set the delivery of the enclosed words only; style prefixes~\citep{tian2026bagpiper}, which set the delivery of the whole line; and director's notes on intent, pacing, or emphasis, each phrased in several templates and registers.
Directing sessions are multi-turn. When a take is not what the director wanted, the next turn gives a note relative to it, such as \dtok{[softer and warmer versus previous]}, and the model produces a new take, with the earlier takes kept in context as audio (Figure~\ref{fig:prompt-direct}).
Multi-speaker dialogue gives each voice one or more reference clips and switches voices in the user text with speaker tokens (Figure~\ref{fig:prompt-clone}).
Whole-scene dubbing describes the characters in text and gives every line up front, joined by supervised separators, so that the delivery of a line can depend on the lines that follow (Figure~\ref{fig:prompt-scene}).
Plain cloning, long-form, infilling, and text-only chat sequences (Appendix~\ref{app:pretrain-formats} and Figure~\ref{fig:prompt-infill}) stay in the mixture so that the pre-training abilities are kept.

\subsection{Reinforcement Learning from Human Preference}\label{sec:rl}

RL starts from the SFT checkpoint and aligns Drama~3's takes with human preference over expressive requests.

\textbf{A single reward model.}
S2 combined three rewards: a transcription-error reward, a speech-quality model, and a speaker-similarity reward.
The transcription-error reward compares what an automatic speech recognition (ASR) system hears with the script, but laughs and long pauses legitimately change what it hears.
The speech-quality model scores the whole utterance at once, so it cannot reward fine-grained control.
The speaker-similarity reward scores a whispered or shouted line lower against a neutral reference.
Each of these rewards either penalizes the expressiveness we want or cannot see it, so Drama~3 scores every take with a single learned model of human preference and uses no other reward.

\textbf{Reward model.}
The reward model (RM) is a smaller speech language model of our own, similar in design to Drama~3 and pre-trained with the same audio tokenizer and sequence format through the end of Stage~2, with a scalar head on the final hidden state of the take's closing token.
It reads exactly the policy's prompt followed by the take's audio tokens, so no tokenizer decoder, ASR, or speaker model runs in the reward path; because it sees the references and the script, voice consistency and content fidelity remain part of what it judges.
It is trained on human preference judgments over pairs of takes with the Bradley--Terry loss~\citep{bradley1952rank}
\begin{equation}
  \mathcal{L}_{\text{RM}} = -\bigl[\, y \log \sigma(s_a - s_b) + (1 - y) \log \sigma(s_b - s_a) \bigr] + 10^{-3} \bigl( s_a^2 + s_b^2 \bigr),
  \label{eq:rm-loss}
\end{equation}
where $s_a, s_b$ are the RM's outputs for takes $a$ and $b$, $\sigma$ is the logistic function, and $y$ is the preference label smoothed~\citep{szegedy2015rethinking} as in conservative direct preference optimization (DPO)~\citep{rafailov2023direct, mitchell2023note}.
Each raw score is standardized by a frozen anchor, clipped, and turned into an expected win rate among the $n$ scored takes sampled for the same prompt:
\begin{equation}
  z_i = \operatorname{clip}\Bigl( \frac{s_i - \mu_0}{\sigma_0}, -3, 3 \Bigr),
  \qquad
  R_i = \frac{1}{n - 1} \sum_{j \neq i} \sigma(z_i - z_j),
  \label{eq:rl-reward}
\end{equation}
where $s_i$ is the RM score of take $i$, and $\mu_0, \sigma_0$ are a mean and a standard deviation fitted once on held-out takes and then kept fixed.
With the anchor fixed, a drift of the whole batch stays visible in the mean $z$, and with the clip, a take gains nothing past three standard deviations.
Here $\sigma(z_i - z_j)$ estimates the probability that take $i$ beats take $j$, and the sum runs over the other $n - 1$ takes of the group, $j \neq i$, so $R_i$ is the rate at which take $i$ beats the rest of its group.
The RM does not score a take with no audio, which stays in its group with reward zero and is left out of this comparison.

\textbf{Sampling.}
For each prompt, the inference engine (Section~\ref{sec:inference}) draws $G = 8$ takes at temperature 1 by plain ancestral sampling, with no top-$p$~\citep{holtzman2020curious} or top-$k$~\citep{fan2018hierarchical} truncation, so that the takes follow exactly the probabilities the engine reports.
Length is limited by a sampling budget rather than a reward.
Each take may run to about 2.5 times the expected duration of its own script, and no take may exceed 4,096 frames, about 190\,s, which bounds runaway takes.

\textbf{Policy optimization.}
We use Group Relative Policy Optimization (GRPO)~\citep{shao2024deepseekmath} with the mean-centered advantages of Dr.~GRPO~\citep{liu2025understanding} and no KL penalty or reference model.
Groups whose reward standard deviation is below 0.05 are dropped, a filter similar to the dynamic sampling of Decoupled Clip and Dynamic Sampling Policy Optimization (DAPO)~\citep{yu2025dapo} but without re-sampling.
With rewards $R_i \in [0, 1]$ (Equation~\ref{eq:rl-reward}),
\begin{equation}
  A_i = R_i - \frac{1}{G} \sum_{j=1}^{G} R_j,
  \qquad
  \mathcal{L}_{\text{RL}}(\theta) = -\frac{1}{|\mathcal{B}|} \sum_{i \in \mathcal{B}} \frac{1}{|y_i|} \sum_{t=1}^{|y_i|} \min\Bigl( \rho_{i,t} A_i,\ \operatorname{clip}\bigl(\rho_{i,t}, 1 - \epsilon_{\text{low}}, 1 + \epsilon_{\text{high}}\bigr) A_i \Bigr),
  \label{eq:rl-loss}
\end{equation}
where $\rho_{i,t} = \pi_\theta(y_{i,t} \mid x_i, y_{i,<t}) / \pi_{\text{eng}}(y_{i,t} \mid x_i, y_{i,<t})$ is the proximal policy optimization (PPO) ratio~\mbox{\citep{schulman2017proximal}}, $x_i$ is the prompt of take $i$, $\pi_\theta$ is the probability that the backbone being trained gives the token, $\pi_{\text{eng}}$ is the probability of the same token reported by the inference engine that sampled it, $\mathcal{B}$ holds the takes of the kept groups, and $y_i$ is the sequence the backbone samples for take $i$, made of the semantic code $q^{(0)}_t$ of every frame and the closing \dtok{<|im_end|>}.
Unlike in S2, the codebook-MTP decoder, which produces the nine acoustic codes of each frame, stays frozen during RL and receives no RL loss.
Because each rollout batch gets one optimizer step, the trainer's weights at the update are the weights that sampled the batch, and the trainer routes each token to the experts that the engine chose, as in rollout routing replay~\citep{ma2025stabilizing}, so $\rho$ departs from 1 only through numerical train--inference mismatch~\citep{qi2025defeating}.
The rollout engines run the routed experts in block-scaled 8-bit floating point (FP8)~\citep{micikevicius2022fp8}, re-quantized after every weight sync, while the trainer runs in bfloat16 (BF16), and this mismatch grows with that precision gap.
We bound this departure with the asymmetric clip of DAPO, $\epsilon_{\text{low}} = 0.2$ and $\epsilon_{\text{high}} = 0.28$.
We keep GRPO's per-take length normalization, which Dr.~GRPO removes.

\textbf{Large batches, few steps.}
Each step uses 1,536 prompts $\times$ $G$ takes, drawn from diverse expressive requests that carry all of their reference clips, at a constant Muon learning rate without weight decay, and the schedule is short.
Without a KL penalty or weight decay, few steps at a small constant learning rate bound how far the policy moves, and large batches average each step over many prompts, so a step follows broad preference rather than the particulars of a few prompts.
In addition, this setup avoids the drift toward reward over-optimization~\citep{gao2023scaling} that small-batch, many-step schedules showed in our RL development.
Recent large language models such as MiMo-V2.6~\citep{xiaomi2026mimo26} likewise use very large RL batches.

\section{Inference Engine}\label{sec:inference}

A streaming request, or stream, plays to its listener while it is generated, so its frames must arrive at least as fast as the codec plays them, about 21.5 per second (Section~\ref{sec:architecture}).
Serving a request has two phases, a prefill that reads its text and reference audio in one pass, and decoding, one step per frame.
One GPU serves many streams at once, in rounds that are each a prefill of new requests or one decode step for all running streams.
We report the per-stream decode rate, the frames per second each stream receives, at a given concurrency, the number of streams a GPU serves at once.

Drama~3 differs from S2 in generation structure, model size, and serving load, and we rewrote the critical path of the inference engine to match.
Its generation structure shortens the sequential path of each frame and reduces the weights it reads (Section~\ref{sec:inference-structure}).
We remove the kernel launches and host waits that CUDA graphs~\citep{gray2019cuda}, which replay a whole step at once, leave in place (Section~\ref{sec:inference-launch}).
The kernels that stay are tuned to the shapes that serving produces (Section~\ref{sec:inference-kernels}), and arriving requests no longer stall the streams already playing (Section~\ref{sec:inference-pd}).

\subsection{Generation Structure}\label{sec:inference-structure}

\begin{figure}[htbp]
  \centering
  \includegraphics[width=\linewidth]{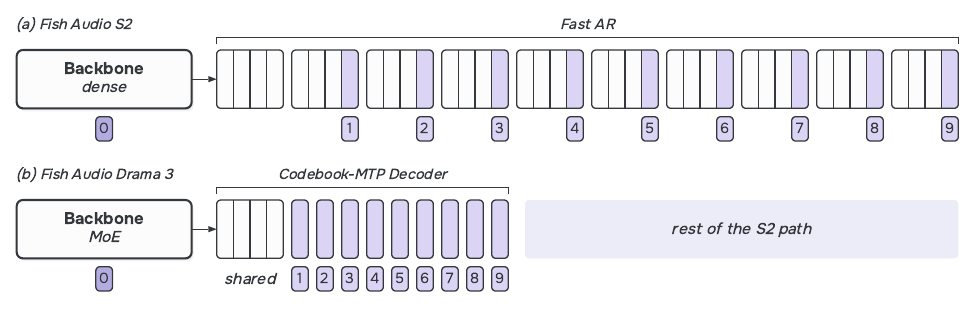}
  \caption{\textbf{Per-frame sequential path after the backbone step.} (a) S2's four-layer Fast AR runs ten passes, one that takes in the backbone's state and one for each of the nine acoustic codes, and reads its weights on every pass. (b) Drama~3's codebook-MTP decoder runs its four shared and nine per-codebook layers once. Boxes are layer passes drawn to scale; each chip is the code emitted by the layer above it.}
  \label{fig:inference-path}
\end{figure}

Streaming batches are usually small, so a step does little arithmetic per weight it reads: decoding is bound by memory bandwidth, and the time of a frame is set mainly by the weights it reads.
In the Dual-AR design of S2, the four-layer Fast AR makes ten passes after the backbone step, so a frame runs 40 layer passes in sequence and reads the Fast AR's weights again on every pass (Figure~\ref{fig:inference-path}a).
At batch size 1, a frame of S2 therefore reads about 15.6\,GB of BF16 weights, two bytes per parameter, the equivalent of about 7.8B parameters, although S2 has about 4.6B.
Drama~3 reads less for two reasons.
First, its codebook-MTP decoder (Section~\ref{sec:codebook-mtp}) runs each of its 13 layers once per frame, so the decoder's weights are read once (Figure~\ref{fig:inference-path}b).
Second, its routed experts are stored in block-scaled FP8, with one scale per $128 \times 128$ block, so each expert parameter takes one byte instead of two.
This is the format in which the RL rollouts already run (Section~\ref{sec:rl}), so serving adds no quantization of its own and runs with the numerics of RL sampling.
A frame of Drama~3 thus reads about 6.0\,GB of weights (Table~\ref{table:inference-structure}), less than two fifths of what S2 reads, with more than seven times as many parameters.

\begin{table}[htbp]
  \centering
  \caption{\textbf{Decoding cost of Drama~3 and S2.} Weights read per frame are computed from the tensor sizes of the checkpoints.}
  \label{table:inference-structure}
  \small
  \setlength{\tabcolsep}{8pt}
  \renewcommand{\arraystretch}{1.12}
  \begin{tabular}{lcc}
    \toprule
    \textbf{Metric} & \textbf{Drama~3} & \textbf{S2} \\
    \midrule
    Layer passes in sequence after the backbone step & \textbf{13} & 40 \\
    Weights read per frame (GB) & \textbf{$\approx$6.0} & $\approx$15.6 \\
    \quad Backbone without routed experts & 2.9 & 7.3 \\
    \quad Eight active routed experts & 1.0 & -- \\
    \quad Codebook-MTP decoder or Fast AR & 2.1 & 8.3 \\
    Layers whose KV cache grows with the context (backbone and decoder) & \textbf{23} & 36 \\
    \bottomrule
  \end{tabular}
\end{table}

At a concurrency of 1, Drama~3 decodes 271 frames per second on one NVIDIA H200 GPU~\citep{nvidia2026h200}, against 177 for S2 measured the same way.
On one NVIDIA B300 GPU~\citep{aubrey2025blackwellultra}, it decodes 295 frames per second; there the experts use the microscaling FP8 format (MXFP8)~\citep{ocp2023mx}, converted from the same FP8 weights.
The one-byte experts matter at larger batches too, where the time of a step still depends mainly on the bytes it moves: on the H200, the FP8 experts raise the per-stream decode rate at a concurrency of 32 by 36\% over BF16 experts.

\subsection{Kernel Launches Below CUDA Graphs}\label{sec:inference-launch}

A step is a sequence of kernels, small programs that the host CPU launches on the GPU, and every kernel costs twice: the host pays to launch it, and the GPU pays a fixed cost to start and drain it.
CUDA graphs record a whole step once and replay it without the host issuing kernels one by one, and programmatic dependent launch (PDL)~\citep{nvidia2026pdl} lets a kernel start while the kernel it depends on is still finishing; neither removes the fixed cost on the GPU.
Before the changes below, we took a census of Drama~3 decoding a single stream on a B300, counting every kernel in a frame with CUDA graphs, PDL, and piecewise prefill graphs (the prefill captured in pieces) enabled.
The median frame took 4.0\,ms and ran 1,171 kernels, with only 0.09\,ms of gaps between them, so the gaps that graphs and PDL target were already closed.
Moving all the data the frame needs would take 0.76\,ms at the GPU's memory bandwidth, one fifth of the frame; the other four fifths went mostly to starting and draining kernels: about 830 kernels moved almost no data yet took 2.7\,ms.
That time shrinks neither with more bandwidth nor under graphs or PDL; the only remedy is fewer kernels.

Most of these 830 kernels come from auxiliary operations that a general-purpose serving framework adds around the computation, such as mask writes, tensor copies, layout changes, and type conversions.
We trace every kernel in the graph back to the framework operator that produced it and decide whether its work is needed, then delete the kernel or fold its work into an adjacent compute kernel.
Fewer kernels is not the goal in itself: the roughly 320 launches per step in the MoE layers are the expert computation and stay.
A kernel that fused the router's general matrix multiplication (GEMM), softmax, and top-8 selection into one launch took 14.5\,$\mu$s, against 7.5\,$\mu$s for the kernels it would have replaced, so it was rejected.
Under PDL, saved kernel time also does not fully reach the wall clock: changes that saved 0.218\,ms of kernel time per step gained 0.14--0.16\,ms in serving.
Every change is therefore accepted on a same-machine A/B comparison, the old and the new engine serving the same load on the same GPU, and Table~\ref{table:inference-launch} lists the accepted changes.

\begin{table}[htbp]
  \centering
  \caption{\textbf{Kernel launches and waits removed below CUDA graphs and PDL.} Gains are in the per-stream decode rate.}
  \label{table:inference-launch}
  \small
  \setlength{\tabcolsep}{4pt}
  \renewcommand{\arraystretch}{1.12}
  \resizebox{\textwidth}{!}{%
  \begin{tabular}{llcl}
    \toprule
    \textbf{Change} & \textbf{Kernels or time} & \textbf{Gain} & \textbf{GPU, concurrency} \\
    \midrule
    Auxiliary decode kernels removed & 1,429 $\to$ 1,278 kernels per step & +1.7\% & B300, 32 \\
    Gated DeltaNet decode fused; sampler counter folded in & $-$100 kernels per step & +4.2\% & B300, 1 \\
    Gated DeltaNet prefill launches merged & 23 $\to$ 13 kernels per layer & +4.4\% & B300, 128 \\
    First-frame sampling captured into CUDA graphs$^{\dagger}$ & 30.9 $\to$ 14.7\,ms host time per round & +15.3\% & H200, 32 \\
    Execution plan built from host-side lengths & 9.75 $\to$ 0.86\,ms GPU idle per round & $\approx$+7\% & B300, 128 \\
    \bottomrule
  \end{tabular}%
  }
  \par\smallskip{\raggedleft\scriptsize $^{\dagger}$Measured on an earlier development model.\par}
\end{table}

\textbf{Auxiliary kernels in decode.}
Every MoE layer launched kernels to write a mask for the padding rows and to copy the shared expert's input, every Gated DeltaNet layer copied the input of its gated normalization into a contiguous layout, and the routing weights were converted to another data type.
Now the routing kernel writes the mask, the copy is made only when the shared expert computes in place, the normalization reads the strided input directly, and the conversion is gone: 1,429 kernels per decode step become 1,278, and the per-stream rate at a concurrency of 32 rises 1.7\%.
The Gated DeltaNet decode, previously three kernels and a gated RMSNorm per layer, is now one kernel that updates the convolution window, applies the gated delta rule, and normalizes, with bit-identical output.
The sampler advances its random-number counter inside the sampling kernel instead of in a launch of its own.
These two changes remove 100 kernels per step at every concurrency, and the per-stream rate at a concurrency of 1 rises 4.2\%; the total per step still varies with concurrency, mainly through the attention kernels, which are chosen by batch size.

\textbf{Auxiliary kernels in prefill.}
The Gated DeltaNet prefill operator launched 23 kernels per layer, most of them copies, reads and writes of per-request state, and index computations.
It now reads the projection tensor by stride, writes the boundary states and initializes the state snapshots in one launch each, and computes its indices once per forward pass rather than in every layer: 13 launches per layer, and the per-stream rate at a concurrency of 128 rises 4.4\%.
Each prefill round also produces the first frame of every new request, through semantic sampling, the 13 decoder layers, and nine codebook samplings, about 100 kernels that the host launched one by one outside the graph.
Capturing this chain into CUDA graphs, together with the Gated DeltaNet prefill chain, cut the host time per prefill round from 30.9 to 14.7\,ms and raised the per-stream rate at a concurrency of 32 by 15.3\% (H200, earlier development model); the Gated DeltaNet prefill operator now runs entirely inside the prefill graph.

\textbf{A wait outside the graph.}
In each prefill round, the attention backend built its execution plan, the schedule of the round's attention work, from indices copied back from the GPU, a copy that waited for all queued backbone work and left the GPU idle.
We now build the plan from lengths that the host already tracks, and the GPU idle time per prefill round at a concurrency of 128 fell from 9.75 to 0.86\,ms, a gain of about 7\% in the per-stream rate.
Removing the wait exposed a data race: the plan's inputs reach the GPU by an asynchronous copy from a pinned host buffer, host memory the GPU reads directly, and the old wait had kept the next round from overwriting that buffer before the copy finished.
Each workspace, the buffers of one round, now has a CUDA event, a marker the GPU records when it has finished reading, that fences the buffer with no waiting in steady state.

\textbf{Rounds that fell out of the graph.}
A CUDA graph helps only a round that matches the form it was captured for.
Every captured graph was guarded by a condition on the batch size or the number of new requests, and a round that failed the condition launched its kernels one by one.
Nine requests arriving at once were enough to cause this fallback.
The scheduler now admits only rounds that the prefill graphs can replay, and the largest captured decode batch equals the admission limit, the most streams the scheduler lets in, so every admitted round runs inside a graph.

\subsection{Shape-Driven Kernel Optimization}\label{sec:inference-kernels}

\begin{figure}[htbp]
  \centering
  \includegraphics[width=0.95\linewidth]{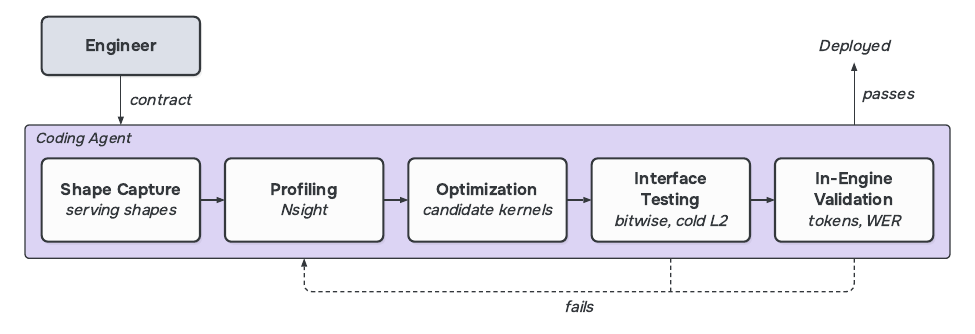}
  \caption{\textbf{Shape-driven kernel optimization loop.} An engineer writes the contract for a round; a coding agent captures the serving shapes, profiles, writes candidate kernels, and tests them, first on the operator interface and then inside the engine. A candidate that fails returns to profiling, and one that passes is deployed. WER: word error rate.}
  \label{fig:inference-loop}
\end{figure}

General-purpose kernel libraries are tuned for generic matrix shapes, the sizes of the matrices a kernel multiplies.
The shapes in TTS serving are set by the workload: decode batches are small and fall into a few sizes, the matrices are thin, the sampler touches only a small slice of the vocabulary, and the Gated DeltaNet geometry is fixed by the model.
A kernel tuned for shapes that serving never produces gains nothing there, so kernel optimization starts from the serving workload and iterates a loop of five steps (Figure~\ref{fig:inference-loop}).

\emph{Shape capture} reads the GEMM dimensions $(M, N, K)$ and batch sizes of every operator from serving logs and graph replays; $M$, the number of rows, is the batch size in decode.
At a concurrency of 32, for example, 91.8\% of decode batches have 31 or 32 rows, so a kernel tuned only for $M \le 16$ gains almost nothing in serving.
\emph{Profiling} attributes the time of each frame to its kernels with the profiler NVIDIA Nsight Systems~\citep{nvidia2026nsightsystems} and measures with Nsight Compute~\citep{nvidia2026nsightcompute} how far each kernel is from the roofline~\citep{williams2009roofline}, the ceiling set by the GPU's memory bandwidth or arithmetic rate.
In a single-operator test at Drama~3's serving dimensions and batch size 256, for example, 97\% of the memory traffic of the Gated DeltaNet decode kernel is the full read and write of the recurrent state, already at 81\% of peak bandwidth.
Faster arithmetic cannot help such a kernel; only fewer bytes can, so on the B300 the state is stored in BF16.
\emph{Optimization} writes candidate kernels from scratch or adapts upstream ones.
\emph{Interface testing} checks the operator alone for bitwise agreement with a reference and times it with a cold L2 cache, one that does not already hold the weights, so that the cache does not flatter small batches.
To keep the cache cold, the test rotates weight copies of at least twice the capacity of the L2, the GPU's on-chip cache, inside the graph.
\emph{In-engine validation} accepts a candidate inside the inference engine on token-level agreement and on the word error rate (WER, Section~\ref{sec:evaluation}) at every concurrency level, and the serving A/B comparison decides deployment.

The last two steps are kept apart because their verdicts often disagree; we recorded nine such cases.
The Gated DeltaNet prefill kernel, for example, was 3.4 times faster as a kernel from the start but lost to the existing one in three serving comparisons, because of graph capture, state movement, and a cold compile at run time.
Once these were resolved, it became the default and raised throughput at a concurrency of 256 by 11.1\%.

Each round of the loop is executed by a coding agent in its own workspace, a separate copy of the code, under a written contract in which an engineer sets the goal, constraints, validation commands, and deployment criteria.
A second agent re-derives each conclusion from the primary evidence, and only candidates that pass every check are deployed.
The FP8 MoE kernels of Drama~3 were developed and iterated in this loop with the Kernel Design Agents workflow~\citep{nvlabs2026kda}.
The FP8 MoE kernel cuts the MoE layer time at $M = 256$ by 19.2\% and raises the per-stream rate in serving at a concurrency of 1 by 3.0\%.

\subsection{Prefill--Decode Disaggregation}\label{sec:inference-pd}

\begin{figure}[htbp]
  \centering
  \includegraphics[width=\linewidth]{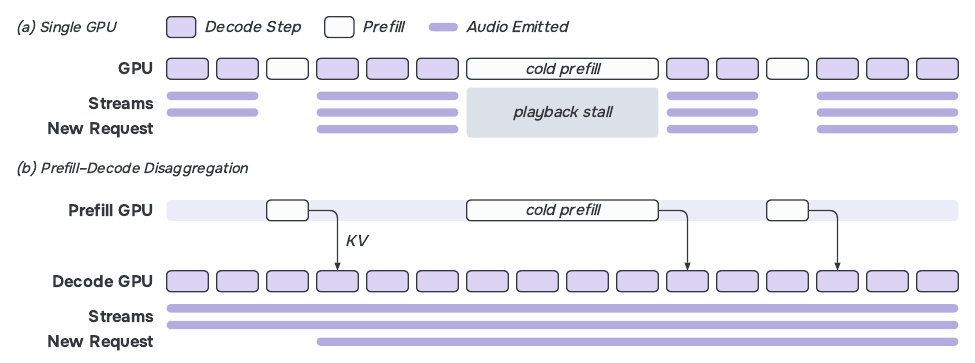}
  \caption{\textbf{Schematic timelines of streaming on one GPU (a) and with prefill--decode disaggregation (b).} On one GPU, every prefill pauses all running streams, and the long prefill of a cold request, one whose reference audio is not cached, causes a playback stall. With disaggregation, the prefill GPU runs the same prefills, each finished prefill writes its KV and recurrent state into the decode GPU, and the streams play without gaps.}
  \label{fig:inference-pd}
\end{figure}

The changes so far make each frame cheaper; this one keeps new requests from interrupting the streams already playing, since decoding that falls behind playback is heard as a stall.
When one GPU serves both phases, every arriving request makes the scheduler insert a prefill round, and during that round no running stream emits a frame (Figure~\ref{fig:inference-pd}a).
On single-GPU workers, these rounds were the main cause of interrupted streams at high concurrency.
A request whose reference audio is already cached needs only a short prefill, but at high request rates such prefills arrive very often, and the longer prefill of a cold request, whose reference audio is not cached, stalls every stream at once.

Prefill--decode disaggregation (PD)~\citep{zhong2024distserve,patel2024splitwise} runs the two phases on different GPUs (Figure~\ref{fig:inference-pd}b).
A PD unit consists of a load balancer, a prefill process, and a decode process.

When a prefill finishes, its KV entries and its Gated DeltaNet states are written within the node straight into the paged KV cache~\citep{kwon2023efficient} and the state buffers of the decode process, in packed transfers with an end-to-end checksum.
The transfer uses NVLink fabric handles on B300 and a PCIe peer-to-peer proxy on NVIDIA RTX PRO 6000 GPUs~\citep{nvidia2026rtxpro6000}.
Each role, prefill or decode, captures only the CUDA graphs it needs, so a unit becomes ready in 47\,s instead of 148\,s.

PD already runs in the production service of Drama~3 on RTX PRO 6000 GPUs.
A capacity test on one RTX PRO 6000 node applies a closed-loop streaming load (each finished stream is replaced by a new one) that reproduces the production mix.
The bar is no playback stall and a per-stream rate of at least 35 frames per second at the first percentile, that is, for all but the slowest one percent of streams.
Under this bar, a unit of one prefill and one decode GPU carries 24 streams per GPU, averaged over both, against about 10 for a single GPU.
In a production canary, a trial on live traffic with PD units and single-GPU workers behind the same router, the PD decode GPUs carry about twice as many streams as the single-GPU workers.
The first percentile of their per-stream rates is 55 frames per second, against 45--47 for the single-GPU workers.
No PD stream falls below the service floor of 42 frames per second, whereas a small share of single-GPU streams does.
The character error rate (CER, Section~\ref{sec:evaluation}) of PD streams is equivalent to that of single-GPU serving, a difference of 0.02 points.

\FloatBarrier

\section{Evaluation}\label{sec:evaluation}

\renewcommand{\topfraction}{0.9}\renewcommand{\bottomfraction}{0.7}\renewcommand{\textfraction}{0.07}\renewcommand{\floatpagefraction}{0.85}
\setcounter{topnumber}{3}\setcounter{bottomnumber}{2}\setcounter{totalnumber}{4}

We evaluate Drama~3 on public benchmarks.
Multilingual cloning is measured on the MiniMax Multilingual Testset~\citep{zhang2025minimaxspeechintrinsiczeroshottexttospeech}.
Editing by infilling is measured on Ming-Freeform-Audio-Edit~\citep{yan2025minguniaudio}, and revision from a note in the next turn on the Step-Audio-Edit benchmark~\citep{yan2025stepaudioeditx}.
All results come from a single Drama~3 model, whereas most baselines are built for one kind of task.
We also analyze what the single preference reward changes (Section~\ref{sec:eval-reward}) and measure cloning from several references and streaming synthesis.

\subsection{Evaluation Setup}\label{sec:eval-protocol}

Drama~3 is run with the prompt formats of Appendix~\ref{app:prompts} and with each benchmark's own reference audio, reference text, and target text, sampling at temperature 0.7 with top-$p$ 0.7 and top-$k$ 30.
Outputs that fail simple signal checks, such as silent output, too little speech, runaway length, or long stretches of noise or silence, are regenerated and rescored.
Baseline numbers are quoted from their papers, except where a table marks a baseline as run by us.
Unless a section or table states otherwise, error rates follow each benchmark's official protocol, with the character error rate (CER) for Chinese, Japanese, and Korean and the word error rate (WER) for the other languages.
Unless stated otherwise, we transcribe Chinese with Paraformer-zh~\citep{gao2023funasr} and the other languages with Whisper-large-v3~\citep{radford2023robust}, and speaker similarity (SIM) is the cosine similarity of the speaker embeddings of WavLM-large~\citep{chen2022wavlm} with an ECAPA-TDNN head~\citep{desplanques2020ecapa}, the speaker model of Seed-TTS-Eval.
Where the conditions of Drama~3 and the baselines differ on a benchmark, we state the difference where the benchmark is introduced.

\subsection{Preference Reward vs.\ Rule-Based Rewards}\label{sec:eval-reward}

Reinforcement learning in Drama~3 optimizes a single learned reward of human preference and uses no transcription or speaker-similarity reward (Section~\ref{sec:rl}).
The rewards it leaves out penalize delivery in the requested style.
In Expresso~\citep{nguyen2023expresso}, each of four speakers reads the same sentences in a default style and in several other styles.
We pair every styled reading with the same speaker's default reading of the same sentence, so the two readings of a pair differ only in style.
With one default-style recording of each speaker as the SIM reference, every style lowers SIM below the default reading, by 0.019 for clearly enunciated reading and by 0.310 for whispering, so a similarity reward prefers the default reading in 62.0\% to 99.7\% of the pairs, depending on the style (Table~\ref{table:expresso}).
Reading with a laugh raises WER by 0.79 points, so a WER reward penalizes laughing.

The reward model of Drama~3 instead responds to the direction.
The voice reference is a default-style clip of the same speaker.
We score both readings of each pair with the reward model twice, once with the style written as a direction in the syntax of Appendix~\ref{app:syntax}, such as \dtok{<whisper>...</whisper>} or \dtok{[laughing]}, and once with the sentence alone.
With the direction, the reward model prefers the whispered reading in 62.2\% of pairs and the laughing reading in 64.4\%, against 10.6\% and 27.1\% without it, whereas a similarity reward prefers the default reading in 99.7\% and 93.9\% of these pairs.
In every style, the direction raises the share of pairs in which the reward model prefers the styled reading (Table~\ref{table:expresso}).

\begin{table}[htbp]
  \centering
  \caption{\textbf{Styled versus default human readings in Expresso.} $\Delta$SIM and $\Delta$WER are means over pairs, styled minus default, $\Delta$WER in percentage points. SIM is to one default-style recording per speaker. SIM prefers default is the share of pairs in which the default reading has the higher SIM. Reward model prefers styled is the share of pairs in which our reward model gives the styled reading the higher score; with direction, the style is written into the script as a direction, and without direction, the script is the sentence alone.}
  \label{table:expresso}
  \small
  \setlength{\tabcolsep}{4pt}
  \renewcommand{\arraystretch}{1.12}
  \resizebox{\textwidth}{!}{%
  \begin{tabular}{lrccccc}
    \toprule
    \multirow{2}{*}{\textbf{Style}} & \multirow{2}{*}{\textbf{Pairs}} & \multirow{2}{*}{$\boldsymbol{\Delta}$\textbf{SIM}} & \multirow{2}{*}{$\boldsymbol{\Delta}$\textbf{WER} (points)} & \multirow{2}{*}{\textbf{SIM prefers default} (\%)} & \multicolumn{2}{c}{\textbf{Reward model prefers styled} (\%)} \\
    \cmidrule(lr){6-7}
    & & & & & With direction & Without direction \\
    \midrule
    Whisper & 1,439 & $-0.310$ & $+0.24$ & $99.7$ & $62.2$ & $10.6$ \\
    Laughing & 1,423 & $-0.193$ & $+0.79$ & $93.9$ & $64.4$ & $27.1$ \\
    Sad & 1,432 & $-0.178$ & $+0.34$ & $96.1$ & $54.3$ & $13.1$ \\
    Confused & 1,372 & $-0.107$ & $-0.61$ & $89.3$ & $26.2$ & $16.3$ \\
    Happy & 1,436 & $-0.063$ & $-0.16$ & $77.9$ & $38.5$ & $29.5$ \\
    Clearly enunciated & 1,447 & $-0.019$ & $-1.06$ & $62.0$ & $8.1$ & $6.6$ \\
    \bottomrule
  \end{tabular}%
  }
\end{table}

A preference reward alone also cuts the failures that rule-based rewards are usually added to catch.
We serve the SFT checkpoint and Drama~3, the policy after RL, through the same inference engine, sampling settings, and tokenizer decoder, and synthesize Seed-TTS-Eval~\citep{anastassiou2024seedttsfamilyhighqualityversatile,bytedance2024seedttseval} with each.
The first take is the output before the regeneration of Section~\ref{sec:eval-protocol}, the final take is the output after it, and a first take is flagged when it fails the signal checks.
Without any transcription reward, RL cuts the share of flagged first takes from 2.82\% to 1.24\% on test-zh and from 3.40\% to 1.84\% on test-en, and the first-take error rates fall with it, from 1.66 to 1.00 CER and from 3.49 to 2.20 WER (Table~\ref{table:sft_rl}).
After the regeneration replaces the flagged takes, Drama~3 still has the lower error rate, 0.83 against 0.84 CER on test-zh and 1.54 against 1.60 WER on test-en, so leaving out the transcription reward costs nothing in intelligibility on the takes that pass the checks either.

\begin{table}[htbp]
  \centering
  \caption{\textbf{SFT checkpoint and Drama~3 on Seed-TTS-Eval.} First take is the output before the regeneration of Section~\ref{sec:eval-protocol}, Final the output after it, and Flagged the share of first takes that fail the signal checks. \textbf{Bold}: better model.}
  \label{table:sft_rl}
  \small
  \setlength{\tabcolsep}{8pt}
  \renewcommand{\arraystretch}{1.12}
  \begin{tabular}{lcccccc}
    \toprule
    \multirow{2}{*}{\textbf{Model}} & \multicolumn{2}{c}{\textbf{test-zh CER} (\%) $\downarrow$} & \multicolumn{2}{c}{\textbf{test-en WER} (\%) $\downarrow$} & \multicolumn{2}{c}{\textbf{Flagged} (\%) $\downarrow$} \\
    \cmidrule(lr){2-3}\cmidrule(lr){4-5}\cmidrule(lr){6-7}
    & First take & Final & First take & Final & test-zh & test-en \\
    \midrule
    SFT checkpoint & 1.66 & 0.84 & 3.49 & 1.60 & 2.82 & 3.40 \\
    \textbf{Fish Audio Drama~3} & \textbf{1.00} & \textbf{0.83} & \textbf{2.20} & \textbf{1.54} & \textbf{1.24} & \textbf{1.84} \\
    \bottomrule
  \end{tabular}
\end{table}

\subsection{Zero-Shot Voice Cloning}\label{sec:eval-cloning}

\subsubsection{Cloning from Multiple References}

Drama~3 can clone one voice from several reference clips, each in its own \dtok{[Reference N]} block (Appendix~\ref{app:prompts}).
No public benchmark reports multi-reference cloning, so we extend the LibriSpeech-PC~\citep{meister2023librispeechpc} test-clean list of F5-TTS~\citep{chen2025f5}, 1,127 sentences from 39 speakers, with two more clips of each target speaker that never include the target sentence.
We score the outputs as OmniVoice~\citep{zhu2026omnivoice} scores itself and the systems it compares on this list (Table~\ref{table:multiref}).
With three references instead of one, SIM to the benchmark's reference clip rises from 0.605 to 0.633.
With one reference, Drama~3's UTMOS~\citep{saeki2022utmos} of 4.31 is above the 4.28 of OmniVoice and CosyVoice~3.

\begin{table}[htbp]
  \centering
  \caption{\textbf{Cloning from one or three references on LibriSpeech-PC test-clean.} The other systems use one reference and are quoted from OmniVoice. Scored as in OmniVoice, with HuBERT-large~\citep{hsu2021hubert} WER, UTMOS~\citep{saeki2022utmos}, and SIM to the first reference. \textbf{Bold}: best per column.}
  \label{table:multiref}
  \small
  \setlength{\tabcolsep}{8pt}
  \renewcommand{\arraystretch}{1.12}
  \begin{tabular}{lccc}
    \toprule
    \textbf{System} & \textbf{WER} (\%) $\downarrow$ & \textbf{SIM} $\uparrow$ & \textbf{UTMOS} $\uparrow$ \\
    \midrule
    Fish Audio Drama~3, one reference & 1.85 & 0.605 & 4.31 \\
    Fish Audio Drama~3, three references & 1.85 & 0.633 & 4.31 \\
    OmniVoice (\citealp{zhu2026omnivoice}) & \textbf{1.30} & \textbf{0.729} & 4.28 \\
    Qwen3-TTS (\citealp{hu2026qwen3ttstechnicalreport}) & 1.60 & 0.704 & \textbf{4.41} \\
    CosyVoice~3 (\citealp{du2025cosyvoice3inthewildspeech}) & 1.59 & 0.694 & 4.28 \\
    \bottomrule
  \end{tabular}
\end{table}

\subsubsection{Multilingual Cloning}

The MiniMax Multilingual Testset~\citep{zhang2025minimaxspeechintrinsiczeroshottexttospeech} covers 24 languages.
We transcribe the outputs with the models of Section~\ref{sec:eval-protocol} and compute the error rates with the public evaluation script of OmniVoice~\citep{zhu2026omnivoice}; ElevenLabs v4, Inworld TTS-2, and Soniox TTS RT v2 are run by us and scored the same way, while OmniVoice, FireRedTTS3-Base~\citep{shen2026fireredtts3}, and Fish Audio S2~\citep{fishaudio2026s2} are quoted from their papers, which used their own scoring.
Table~\ref{table:minimax_err} reports the error rates.
Drama~3's error rate is lower than that of OmniVoice in 15 languages, among them Chinese, English, French, German, Korean, and Thai, and lower than that of FireRedTTS3-Base in 17.
It has a lower error rate than OmniVoice, FireRedTTS3-Base, and Fish Audio S2 in eleven languages: Arabic, Chinese, Dutch, English, Finnish, French, German, Polish, Portuguese, Turkish, and Vietnamese.
Against the systems we run ourselves, it is lower than Inworld TTS-2 in 18 languages, than Soniox TTS RT v2 in 21 of the 23 it supports, and than ElevenLabs v4 in Chinese, Dutch, English, German, Korean, Russian, Spanish, Thai, and Turkish; in Chinese, Dutch, English, German, and Turkish it is lower than every other system in the table.

\begin{table}[htbp]
  \centering
  \caption{\textbf{MiniMax Multilingual Testset, error rate (\%).} ElevenLabs v4~\citep{elevenlabs2026v4}, Inworld TTS-2~\citep{inworld2026tts2}, and Soniox TTS RT v2~\citep{soniox2026ttsrtv2} are run by us through their public APIs on the same items. OmniVoice, FireRedTTS3-Base, and Fish Audio S2 are quoted from their papers. \textbf{Bold}: lowest per row.}
  \label{table:minimax_err}
  \small
  \setlength{\tabcolsep}{3pt}
  \renewcommand{\arraystretch}{1.05}
  \resizebox{\textwidth}{!}{%
  \begin{tabular}{llccccccc}
    \toprule
    \textbf{Language} & \textbf{Metric} & \textbf{Drama~3} & OmniVoice & FireRedTTS3-Base & Fish Audio S2 & ElevenLabs v4 & Inworld TTS-2 & Soniox TTS RT v2 \\
    \midrule
    Arabic & CER & 0.76 & 1.39 & 1.75 & 3.50 & \textbf{0.71} & 1.39 & 2.30 \\
    Cantonese & CER & 22.70 & \textbf{17.71} & 40.32 & 30.67 & 19.74 & 24.14 & -- \\
    Chinese & CER & \textbf{0.63} & 1.01 & 0.91 & 0.73 & 0.66 & 0.97 & 1.32 \\
    Czech & WER & 3.53 & 2.86 & 3.17 & 2.84 & \textbf{2.13} & 3.95 & 8.36 \\
    Dutch & WER & \textbf{0.59} & 1.36 & 1.15 & 0.99 & 1.00 & 1.53 & 2.01 \\
    English & WER & \textbf{1.13} & 1.56 & 2.12 & 1.62 & 1.56 & 1.56 & 1.94 \\
    Finnish & WER & 3.09 & 3.75 & 3.10 & 3.33 & \textbf{2.43} & 4.74 & 4.14 \\
    French & WER & 2.68 & 3.35 & 5.28 & 3.05 & \textbf{2.63} & \textbf{2.63} & 4.12 \\
    German & WER & \textbf{0.48} & 0.96 & 0.69 & 0.55 & 0.90 & 0.83 & 1.10 \\
    Greek & CER & 3.39 & 1.06 & 1.24 & 5.74 & \textbf{0.56} & 0.98 & 1.33 \\
    Hindi & CER & 6.35 & \textbf{4.33} & 7.02 & 14.64 & 5.78 & 7.16 & 8.76 \\
    Indonesian & WER & 4.93 & 1.97 & \textbf{1.42} & 1.46 & 1.59 & 1.67 & 5.08 \\
    Italian & WER & 1.52 & 2.07 & 2.28 & \textbf{1.27} & 1.29 & 1.76 & 2.31 \\
    Japanese & CER & 5.02 & 4.03 & 3.60 & \textbf{2.76} & 3.66 & 4.23 & 5.36 \\
    Korean & CER & 2.36 & 2.65 & 2.42 & \textbf{1.18} & 2.63 & 2.65 & 3.43 \\
    Polish & WER & 0.82 & 0.87 & 1.22 & 1.26 & \textbf{0.77} & 1.39 & 2.62 \\
    Portuguese & WER & \textbf{1.11} & 2.51 & 1.79 & 1.14 & \textbf{1.11} & 1.79 & 1.40 \\
    Romanian & WER & 4.94 & 2.42 & 1.93 & 10.74 & \textbf{1.60} & 2.86 & 4.63 \\
    Russian & WER & \textbf{2.23} & \textbf{2.23} & 3.28 & 2.40 & 2.32 & 2.65 & 4.05 \\
    Spanish & WER & 0.96 & 1.03 & 1.21 & 0.91 & 1.09 & \textbf{0.90} & 1.67 \\
    Thai & CER & 2.27 & 3.98 & \textbf{1.87} & 4.23 & 3.08 & 4.19 & 5.47 \\
    Turkish & WER & \textbf{0.52} & 2.17 & 0.92 & 0.87 & 0.87 & 1.82 & 1.91 \\
    Ukrainian & WER & 2.10 & 1.77 & \textbf{0.55} & 2.30 & 1.37 & 2.18 & 3.06 \\
    Vietnamese & CER & 0.80 & 1.37 & 0.86 & 7.41 & \textbf{0.61} & 1.25 & 3.36 \\
    \bottomrule
  \end{tabular}%
  }
\end{table}

\subsection{Streaming Synthesis}\label{sec:eval-streaming}

Each sentence is synthesized on its own in two modes: streamed, where its audio is sent out chunk by chunk while the rest is still being generated, and non-streamed, where the whole take is sent out once it is finished.
We do this for all 1,088 Seed-TTS-Eval test-en sentences.
Streaming costs no quality: on the same sentences, streamed takes reach a WER of 1.86 and non-streamed takes 1.83, and both reach a SIM of 0.640.

\subsection{Editing by Infilling}\label{sec:eval-editing}

Drama~3 edits a recording by infilling~\citep{le2023voicebox,peng2024voicecraft}.
We align the source audio to its transcript with Qwen3-ForcedAligner-0.6B~\citep{shi2026qwen3asr}.
The edit window runs from the first to the last word that differs between the source and target texts, with characters as words in Chinese.
A deletion leaves no new text, so its window also covers the unchanged word before it, or after it at the start of the clip.
Drama~3 regenerates the audio inside the window with the infilling prompt of Figure~\ref{fig:prompt-infill}, which holds the full target text and the audio tokens of the source before and after the window.
The whole token sequence, with the new tokens inside the window, is then decoded to audio.

Ming-Freeform-Audio-Edit, which we shorten to Ming-Freeform, asks for deletions, insertions, and substitutions in English and Chinese.
Each item gives a source recording, its transcript, an edit instruction, and the target text, and Drama~3 edits from the target text.
The basic split has instructions of limited variety, and the larger full split, which the benchmark's evaluation script calls open, has more natural and varied instructions and more substitutions.
We score the outputs with the protocol of CosyEdit~\citep{chen2026cosyedit} and CosyEdit2~\citep{chen2026cosyedit2}, which reports WER, SIM against the source recording, and MAE$_\text{D}$.
MAE$_\text{D}$ is the mean absolute difference between the DNSMOS P.835 overall score~\citep{reddy2021dnsmosp835} of the edited clip and that of its source, and it is small when the edit leaves the perceived quality of the recording unchanged.
Table~\ref{table:ming_cosyedit} compares Drama~3 on the basic split with VoiceCraft-X~\citep{zheng2025voicecraftx}, LEMAS-Edit~\citep{zhao2026lemas}, ECPA~\citep{ren2026ecpa}, SSR-Speech~\citep{wang2025ssrspeech}, and CosyEdit2, as reported by CosyEdit2.
Among them, SSR-Speech reconstructs the unedited spans inside its decoder, which takes the source waveform as an additional input, and VoiceCraft-X decodes the source tokens together with the generated ones; the Drama~3 rows here are likewise decoded whole, without pasting the source recording back.
In Chinese, Drama~3 has the highest SIM on insertion and substitution, 0.909 and 0.898 against 0.89 for CosyEdit2, the lowest WER on deletion, 7.82 against 8.23, and the smallest MAE$_\text{D}$ on deletion and substitution.

\begin{table}[htbp]
  \centering
  \caption{\textbf{Ming-Freeform basic split under the CosyEdit protocol.} WER in \%. Baselines are quoted from CosyEdit2; -- marks a metric not reported. \textbf{Bold}: best per column within each language.}
  \label{table:ming_cosyedit}
  \small
  \setlength{\tabcolsep}{3pt}
  \renewcommand{\arraystretch}{1.1}
  \resizebox{\textwidth}{!}{%
  \begin{tabular}{lccccccccc}
    \toprule
    \multirow{2}{*}{\textbf{System}} & \multicolumn{3}{c}{\textbf{Insertion}} & \multicolumn{3}{c}{\textbf{Deletion}} & \multicolumn{3}{c}{\textbf{Substitution}} \\
    \cmidrule(lr){2-4}\cmidrule(lr){5-7}\cmidrule(lr){8-10}
    & WER $\downarrow$ & SIM $\uparrow$ & MAE$_\text{D}$ $\downarrow$ & WER $\downarrow$ & SIM $\uparrow$ & MAE$_\text{D}$ $\downarrow$ & WER $\downarrow$ & SIM $\uparrow$ & MAE$_\text{D}$ $\downarrow$ \\
    \midrule
    \multicolumn{10}{l}{\textit{English}} \\
    \textbf{Fish Audio Drama~3} & 3.03 & 0.905 & 0.208 & 5.79 & 0.838 & 0.174 & 2.08 & 0.866 & 0.181 \\
    VoiceCraft-X (\citealp{zheng2025voicecraftx}) & 5.74 & 0.85 & 0.166 & 9.71 & 0.80 & 0.172 & 7.29 & 0.81 & 0.155 \\
    LEMAS-Edit (\citealp{zhao2026lemas}) & 6.04 & 0.81 & 0.158 & 9.29 & 0.80 & 0.167 & 6.09 & 0.80 & 0.167 \\
    ECPA (\citealp{ren2026ecpa}) & 4.50 & 0.82 & -- & 6.91 & 0.77 & -- & 4.13 & 0.78 & -- \\
    SSR-Speech (\citealp{wang2025ssrspeech}) & \textbf{1.75} & \textbf{0.94} & 0.139 & \textbf{5.22} & \textbf{0.91} & 0.132 & 1.90 & \textbf{0.89} & 0.146 \\
    CosyEdit2 (\citealp{chen2026cosyedit2}) & 1.90 & 0.93 & \textbf{0.107} & 5.52 & 0.90 & \textbf{0.131} & \textbf{1.43} & \textbf{0.89} & \textbf{0.137} \\
    \midrule
    \multicolumn{10}{l}{\textit{Chinese}} \\
    \textbf{Fish Audio Drama~3} & 2.15 & \textbf{0.909} & \textbf{0.113} & \textbf{7.82} & 0.827 & \textbf{0.105} & 1.87 & \textbf{0.898} & \textbf{0.100} \\
    VoiceCraft-X (\citealp{zheng2025voicecraftx}) & 3.77 & 0.84 & 0.128 & 13.62 & 0.77 & 0.203 & 4.21 & 0.82 & 0.158 \\
    LEMAS-Edit (\citealp{zhao2026lemas}) & 6.21 & 0.81 & 0.130 & 17.43 & 0.75 & 0.159 & 8.16 & 0.82 & 0.157 \\
    CosyEdit2 (\citealp{chen2026cosyedit2}) & \textbf{1.38} & 0.89 & \textbf{0.113} & 8.23 & \textbf{0.85} & 0.166 & \textbf{1.26} & 0.89 & 0.110 \\
    \bottomrule
  \end{tabular}%
  }
\end{table}

We also run Cartesia Sonic-3~\citep{cartesia2026api} through its infill API, which takes the audio before and after a window and the new text of the window and returns speech for the window.
Cartesia gets the edit windows of Drama~3 and a voice cloned from the source.
Because its API returns only the audio of the window, the new audio of both systems is pasted into the source recording here, matched in loudness, with a 20~ms equal-power crossfade at each joint (Table~\ref{table:ming_cartesia}).
Drama~3 has the lower WER on both splits in both languages, and in Chinese its WER is less than half of Cartesia's, 3.69 against 8.64 on the basic split and 2.35 against 6.53 on the full split.

\begin{table}[htbp]
  \centering
  \caption{\textbf{Ming-Freeform WER (\%) against Cartesia Sonic-3.} Each cell averages deletion, insertion, and substitution, with the new audio pasted into the source recording. Cartesia is run by us through its public API. \textbf{Bold}: lowest per column.}
  \label{table:ming_cartesia}
  \small
  \setlength{\tabcolsep}{8pt}
  \renewcommand{\arraystretch}{1.12}
  \begin{tabular}{lcccc}
    \toprule
    \multirow{2}{*}{\textbf{System}} & \multicolumn{2}{c}{\textbf{Basic}} & \multicolumn{2}{c}{\textbf{Full}} \\
    \cmidrule(lr){2-3}\cmidrule(lr){4-5}
    & English & Chinese & English & Chinese \\
    \midrule
    \textbf{Fish Audio Drama~3} & \textbf{3.49} & \textbf{3.69} & \textbf{3.43} & \textbf{2.35} \\
    Cartesia Sonic-3, infill (\citealp{cartesia2026api}) & 4.58 & 8.64 & 4.68 & 6.53 \\
    \bottomrule
  \end{tabular}
\end{table}

\subsection{Multi-Turn Revision}\label{sec:eval-revision}

The Step-Audio-Edit benchmark~\citep{yan2025stepaudioeditx} clones a voice and then asks for up to three rounds of revision toward a target emotion or speaking style, or for one round that adds a paralinguistic event, such as a laugh or a breath, at a marked position.
Iteration 0 is plain cloning, and each later iteration revises the most recent take.
Gemini-2.5-Pro~\citep{comanici2025gemini} judges every take with the official prompts, and we score its answers with the official scripts.
For emotion and style, the judge names the one emotion or style it hears, and the accuracy is the share of takes named with their target.
The emotion labels are the five targets and a neutral one.
For style, the official script counts only the takes named with one of the seven target styles.
Where the official script has no valid takes for a speaker and a style, that pair is left out of the average.
For the paralinguistic event, the judge gives a score from 1 to 3, where 1 means the event is missing or misplaced, 2 means it is present with a flaw such as a faint sound or an audible join, and 3 means it is clear, in place, and seamless.
Drama~3 revises with a note in the next turn, such as \dtok{[happier versus previous]}, and keeps the original reference and the most recent take in context, which corresponds to the Prompt-Fixed setting of the Step-Audio-EditX paper, so we compare with Step-Audio-EditX in that setting.
Paralinguistic events are added by inserting a point tag such as \dtok{[coughs]} at the target position and infilling the audio there.

Table~\ref{table:stepaudioedit} shows that every revision round raises the emotion and style accuracy of Drama~3, from 55.5 to 58.2 on emotion and from 76.5 to 81.8 on style over three rounds.
Adding the event with a point tag and infilling lifts its paralinguistic score from 1.32 to 2.32.

\begin{table}[htbp]
  \centering
  \caption{\textbf{Step-Audio-Edit benchmark.} Step-Audio-EditX is quoted from the Prompt-Fixed setting of its paper, with the judge scores reported there; Drama~3 is judged by our own run of the official Gemini-2.5-Pro protocol. Emotion and style are accuracies (\%). \textbf{Bold}: better system.}
  \label{table:stepaudioedit}
  \small
  \setlength{\tabcolsep}{8pt}
  \renewcommand{\arraystretch}{1.12}
  \begin{tabular}{llcccc}
    \toprule
    \textbf{Metric} & \textbf{System} & \textbf{Iter.~0} & \textbf{Iter.~1} & \textbf{Iter.~2} & \textbf{Iter.~3} \\
    \midrule
    \multirow{2}{*}{Emotion $\uparrow$} & \textbf{Fish Audio Drama~3} & \textbf{55.5} & 57.2 & 57.5 & 58.2 \\
    & Step-Audio-EditX (\citealp{yan2025stepaudioeditx}) & 53.6 & \textbf{66.8} & \textbf{68.7} & \textbf{69.3} \\
    \midrule
    \multirow{2}{*}{Style $\uparrow$} & \textbf{Fish Audio Drama~3} & \textbf{76.5} & \textbf{79.4} & \textbf{80.5} & \textbf{81.8} \\
    & Step-Audio-EditX (\citealp{yan2025stepaudioeditx}) & 45.6 & 62.7 & 64.2 & 63.8 \\
    \midrule
    \multirow{2}{*}{Paralinguistic $\uparrow$} & \textbf{Fish Audio Drama~3} & 1.32 & 2.32 & -- & -- \\
    & Step-Audio-EditX (\citealp{yan2025stepaudioeditx}) & \textbf{1.91} & \textbf{2.89} & -- & -- \\
    \bottomrule
  \end{tabular}
\end{table}

\FloatBarrier

\section{Conclusion}\label{sec:conclusion}

In this report, we present Fish~Audio~Drama~3, a speech generation model that turns text into directed performances. It clones voices from multiple references, follows natural-language direction, revises a take when the next turn gives a note, performs multi-character scenes, edits existing speech by infilling, and synthesizes long-form and streaming audio.
These capabilities rest on three design choices.
A hybrid linear-attention mixture-of-experts backbone is paired with a codebook multi-token-prediction decoder (Section~\ref{sec:architecture}) whose layers attend over the whole sequence, so acoustic detail is predicted with direct access to the reference audio and all previous frames.
Multi-turn revision, scene dubbing, editing by infilling, and streaming text input are supported by one model, each written as a conversation over the same token stream (Section~\ref{sec:conv-format}, \emph{Conversation format}).
Reinforcement learning (Section~\ref{sec:rl}) uses a single learned human-preference reward and no rule-based rewards, which would penalize laughs, pauses, and whispers.
Serving these capabilities is the work of the inference engine (Section~\ref{sec:inference}), which runs the codebook-MTP decoder once per frame and the experts in the block-scaled FP8 of the RL rollouts, removes the kernel launches that CUDA graphs leave in place, optimizes kernels on the shapes of live serving, and separates prefill from decode so that new requests do not stall running streams.
On one H200, Drama~3 decodes a single stream at 271 frames per second against 177 for S2, with more than seven times as many parameters.
We evaluate this single model on voice cloning, streaming synthesis, editing by infilling, and multi-turn revision, mostly on public benchmarks (Section~\ref{sec:evaluation}).
We hope Fish~Audio~Drama~3 serves as an open foundation for expressive, directable speech generation.

\section{Authors}

\textbf{Core Contributors}

The core contributors are listed in alphabetical order by last name:

Dawei Han, Shidong Li, Shijia Liao, Jiahua Liu, Songting Liu, Qingzheng Wang, Yuxuan Wang

\textbf{Contributors}

The contributors are listed in alphabetical order by last name:

Xin Chen, Yifan Cheng, Tianyu Li, Gangin Park, Xuehao Yang, Chang Zeng, Haoyang Zhang, Ruoyi Zhang, Xiangyu Zhang

For correspondence, please contact oss@fish.audio.

\bibliography{drama3/drama3}
\bibliographystyle{colm2024_conference}

\clearpage
\appendix

\section{Pre-training Sequence Formats}\label{app:pretrain-formats}

Figures~\ref{fig:pre-tts}--\ref{fig:pre-stream} show the TTS conversation of Stages~1 and~2 and the long-form, line-separated, and streaming sequences of Stage~3; text documents are plain text.
\texttt{\{\{...\}\}} stands for a span of audio frames, and long lines are broken at spaces for display.
\dtok{<|voice|>} is the modality token that opens an audio turn.
The system instructions are the ones in use; all other text is illustrative.
In every sequence, loss falls only on the audio frames of the assistant turns, on the separators of the line-separated form, and on the closing \dtok{<|im_end|>}.
Speech infilling uses the layout of Figure~\ref{fig:prompt-infill}; its loss covers either the whole middle or only its first and last few frames and the closing token, so the seams are always supervised.

\begin{figure}[htbp]
  \centering
  \begin{minipage}{\linewidth}
  \hrule\vspace{2pt}
  \footnotesize
\begin{verbatim}
<|im_start|>system
convert the provided text to speech reference to the following:

Text:
<|speaker:0|> I told you already, the train leaves at seven.

Speech:
{{reference clip}}<|im_end|>
<|im_start|>user
<|speaker:0|> [sighs] Fine. If we miss it, it's on you. [audio tokens: 60] [language: en]<|im_end|>
<|im_start|>assistant
<|voice|>{{utterance 1}}<|im_end|>
<|im_start|>user
<|speaker:0|> Seven sharp. I'll be at the door.<|im_end|>
<|im_start|>assistant
<|voice|>{{utterance 2}}<|im_end|>
\end{verbatim}
  \vspace{2pt}\hrule
  \end{minipage}
  \caption{\textbf{Stages 1--2: a multi-turn TTS conversation.}}
  \label{fig:pre-tts}
\end{figure}

The sequence of Figure~\ref{fig:pre-tts} teaches plain synthesis and cloning. Consecutive utterances of one recording follow each other as user and assistant turns, with any reference clips in the system turn (Section~\ref{sec:conv-format}, \emph{Conversation format}), and the user text may carry point tags and language and duration hints.

\begin{figure}[htbp]
  \centering
  \begin{minipage}{\linewidth}
  \hrule\vspace{2pt}
  \footnotesize
\begin{verbatim}
<|im_start|>system
convert the provided text to speech reference to the following:

Text:
{{transcript of a clip from later in the recording}}

Speech:
{{that clip}}<|im_end|>
<|im_start|>user
<|speaker:0|> {{transcript of the recording up to the clip}}<|im_end|>
<|im_start|>assistant
<|voice|>{{the recording up to the clip}}<|im_end|>
\end{verbatim}
  \vspace{2pt}\hrule
  \end{minipage}
  \caption{\textbf{Stage 3: long-form synthesis.} The reference clip is cut from later in the same recording.}
  \label{fig:pre-longform}
\end{figure}

The sequence of Figure~\ref{fig:pre-longform} voices a long recording from its transcript in one turn.
Its reference clips are cut from later in the same recording and the target stops where the first clip begins, so no audio is both context and target.
With several clips, the system turn either concatenates them as in S2 or gives each its own numbered block, at random.
Some of these sequences have no reference; the system turn then holds only the instruction and the loss covers the whole recording.

\begin{figure}[htbp]
  \centering
  \begin{minipage}{\linewidth}
  \hrule\vspace{2pt}
  \footnotesize
\begin{verbatim}
<|im_start|>system
convert the provided text to speech reference to the following:

Text:
{{transcript of a clip from later in the recording}}

Speech:
{{that clip}}

convert the provided text to speech; the lines are separated by <|audio_separator|>
— synthesize each line in order and output <|audio_separator|> between adjacent audio
segments<|im_end|>
<|im_start|>user
It was late.
<|audio_separator|>
The hall was empty.
<|audio_separator|>
Nobody answered.<|im_end|>
<|im_start|>assistant
<|voice|>{{line 1}}
<|audio_separator|>
{{line 2}}
<|audio_separator|>
{{line 3}}<|im_end|>
\end{verbatim}
  \vspace{2pt}\hrule
  \end{minipage}
  \caption{\textbf{Stage 3: line-separated synthesis.} The reference clip precedes the line instruction in the system turn.}
  \label{fig:pre-lines}
\end{figure}

The sequence of Figure~\ref{fig:pre-lines} voices one text line by line. The separator \dtok{<|audio_separator|>} joins the lines in the user turn, and the model emits one audio segment per line with the separator between segments; its reference clips are chosen as in Figure~\ref{fig:pre-longform}, and some of these sequences carry none.
The separators are supervised outputs, so the model learns where segments end and how many there are.
The separators act as anchors between the text and the audio.
If the model skips or repeats a sentence inside one segment, it still starts the next segment from the next line, so the mistake does not spread; and if the number of segments differs from the number of lines, the run-on is detected.

\begin{figure}[htbp]
  \centering
  \begin{minipage}{\linewidth}
  \hrule\vspace{2pt}
  \footnotesize
\begin{verbatim}
<|im_start|>system
streaming speech continuation with one-chunk lookahead split by <|endofturn|><|im_end|>
<|im_start|>user
It was late<|endofturn|>and the hall was empty.<|im_end|>
<|im_start|>assistant
<|voice|>{{"It was late"}}<|im_end|>
<|im_start|>user
and the hall was empty.<|endofturn|>Nobody answered.<|im_end|>
<|im_start|>assistant
<|voice|>{{"and the hall was empty."}}<|im_end|>
<|im_start|>user
Nobody answered.<|endofturn|><|im_end|>
<|im_start|>assistant
<|voice|>{{"Nobody answered."}}<|im_end|>
\end{verbatim}
  \vspace{2pt}\hrule
  \end{minipage}
  \caption{\textbf{Stage 3: streaming interleave.}}
  \label{fig:pre-stream}
\end{figure}

The sequence of Figure~\ref{fig:pre-stream} is for text that arrives in pieces.
Each user turn holds the chunk to voice and, after \dtok{<|endofturn|>}, the chunk that follows, so the model knows what comes next while voicing the current chunk; an empty next chunk marks the last round.
Chunks are consecutive spans of the transcript.

\section{Prompt Examples}\label{app:prompts}

Figures~\ref{fig:prompt-clone}--\ref{fig:prompt-infill} render the main prompt formats in full.
As in Appendix~\ref{app:pretrain-formats}, \texttt{\{\{...\}\}} stands for a span of audio frames. In the backbone stream, each frame occupies one position, entered as the fused embedding of its codes 0 and 1; codes 2--9 enter the decoder as side input (Section~\ref{sec:partial-fusion}).
The system instructions of the cloning, directing, and infilling prompts are shown exactly as used; all other text is illustrative, and long lines are broken at spaces for display.
Loss applies only to assistant turns (Section~\ref{sec:conv-format}, \emph{Conversation format}), in a spoken turn to all ten codebooks of every supervised frame and to the closing \texttt{<|im\_end|>}; the infilling format additionally weights its loss toward the seams (Appendix~\ref{app:pretrain-formats}).
At inference, the prompt ends with the last \texttt{<|voice|>}, and the model generates frames until it emits \texttt{<|im\_end|>}.

\begin{figure}[htbp]
  \centering
  \begin{minipage}{\linewidth}
  \hrule\vspace{2pt}
  \footnotesize
\begin{verbatim}
<|im_start|>system
convert the provided text to speech reference to the following:

[Reference 1]
Text: <|speaker:0|> I told you already, the train leaves at seven.
Speech: {{voice A, clip 1}}

[Reference 2]
Text: <|speaker:0|> Fine. But if we miss it, that's on you.
Speech: {{voice A, clip 2}}

[Reference 3]
Text: <|speaker:1|> Seven? Nobody said anything about seven.
Speech: {{voice B}}<|im_end|>
<|im_start|>user
<|speaker:0|> [whispering] Don't wake the baby.
<|audio_separator|>
<|speaker:1|> [laughing] Okay, okay, I'm going.<|im_end|>
<|im_start|>assistant
<|voice|>{{turn 1}}<|audio_separator|>{{turn 2}}<|im_end|>
\end{verbatim}
  \vspace{2pt}\hrule
  \end{minipage}
  \caption{\textbf{Cloning a two-speaker dialogue from three reference clips, two of one voice.}}
  \label{fig:prompt-clone}
\end{figure}

Figure~\ref{fig:prompt-clone} casts two voices from three reference clips.
Speaker tokens are global to the request (Section~\ref{sec:conv-format}, \emph{Conversation format}). The identity in \texttt{<|speaker:S|>} is independent of the block number, so References 1 and 2, two clips of one voice, carry the same speaker token, and Reference 3 introduces the second voice.

\begin{figure}[htbp]
  \centering
  \begin{minipage}{\linewidth}
  \hrule\vspace{2pt}
  \footnotesize
\begin{verbatim}
<|im_start|>system
convert the provided text to speech reference to the following:

[Reference 1]
Text: <|speaker:0|> Honestly, I thought we had a real shot this year.
Speech: {{reference clip}}<|im_end|>
<|im_start|>user
<|speaker:0|> [with heavy, deflated sadness] The results just came in. We didn't make it.<|im_end|>
<|im_start|>assistant
<|voice|>{{take 1}}<|im_end|>
<|im_start|>user
<|speaker:0|> [softer and warmer versus previous] The results just came in. We didn't make it.<|im_end|>
<|im_start|>assistant
<|voice|>{{take 2}}<|im_end|>
<|im_start|>user
<|speaker:0|> Direction: more amused than before; drop the hurt.
Line: The results just came in. We didn't make it.<|im_end|>
<|im_start|>assistant
<|voice|>{{take 3}}<|im_end|>
\end{verbatim}
  \vspace{2pt}\hrule
  \end{minipage}
  \caption{\textbf{Multi-turn directing, in which each later turn gives a note on the previous take and the model produces a new one.}}
  \label{fig:prompt-direct}
\end{figure}

In Figure~\ref{fig:prompt-direct}, turn~1 gives an absolute direction as a style prefix, turn~2 a relative note as a bracketed prefix, and turn~3 a relative note in the director's-note register.
At inference, the earlier takes are the audio that the model generated in previous turns, so each note is interpreted against what the listener actually heard.

\begin{figure}[htbp]
  \centering
  \begin{minipage}{\linewidth}
  \hrule\vspace{2pt}
  \footnotesize
\begin{verbatim}
<|im_start|>system
Scene: a small office on a Monday morning; the weekly meeting is about to start.
Characters:
- Mr. Carter: steady male voice, about 40, formal.
- Emma: young female voice, sleepy, half talking to herself.
Voice every line of the script in order, with <|audio_separator|> between consecutive lines.<|im_end|>
<|im_start|>user
Mr. Carter: [clears throat][serious] Good morning, everyone. Let's get started.
<|audio_separator|>
Mr. Carter: [reading from a sheet] First item: the launch date.
<|audio_separator|>
Emma: [softly][to herself] Here we go again...<|im_end|>
<|im_start|>assistant
<|voice|>{{line 1}}<|audio_separator|>{{line 2}}<|audio_separator|>{{line 3}}<|im_end|>
\end{verbatim}
  \vspace{2pt}\hrule
  \end{minipage}
  \caption{\textbf{Multi-character scene dubbing from text descriptions.}}
  \label{fig:prompt-scene}
\end{figure}

In Figure~\ref{fig:prompt-scene}, whose content is illustrative, the whole scene is one turn and every line is given up front.
Characters are cast from their text descriptions rather than from reference audio, each line starts with its speaker's name, and the separators are supervised outputs, so the model learns to emit exactly one audio segment per line, also when one character speaks consecutive lines (lines 1 and 2).
The line-separated long-form format uses the same layout, with the lines of one text joined by separators in the user turn and one audio segment per line with supervised separators in the output. Only the scene and character descriptions are absent.

\begin{figure}[htbp]
  \centering
  \begin{minipage}{\linewidth}
  \hrule\vspace{2pt}
  \footnotesize
\begin{verbatim}
<|im_start|>system
fill in the missing part of the speech; the suffix follows <|fim_suffix|> and the prefix follows
<|fim_prefix|><|im_end|>
<|im_start|>user
<|speaker:0|> The demo is Thursday, not Friday. [audio tokens: 14]
<|fim_suffix|>{{audio of ", not Friday."}}<|fim_prefix|>{{audio of "The demo is"}}<|im_end|>
<|im_start|>assistant
<|voice|>{{new audio of "Thursday"}}<|im_end|>
\end{verbatim}
  \vspace{2pt}\hrule
  \end{minipage}
  \caption{\textbf{Speech editing by infilling.}}
  \label{fig:prompt-infill}
\end{figure}

Figure~\ref{fig:prompt-infill} edits a recording by replacing one spoken word (``Wednesday'' becomes ``Thursday'').
The prefix audio is the original recording before the edited word and the suffix audio is the recording after it, while the user turn carries the full new transcript, including the replacement word.
Following the suffix-prefix-middle ordering of fill-in-the-middle training~\citep{bavarian2022efficient}, the suffix comes before the prefix, so the prefix sits directly next to the generated audio and the model continues from it; the optional duration hint gives the length of the new span in frames.

\section{Control Syntax Reference}\label{app:syntax}

Table~\ref{table:syntax} lists the controls that Drama~3 reads in its prompts.
Structural tokens, written \texttt{<|name|>}, are single vocabulary entries; every other control is ordinary text.
Section~\ref{sec:training} names the training stage at which each format is introduced, and Appendix~\ref{app:prompts} shows complete prompts.

\providecommand{\ph}[1]{\textnormal{\textit{#1}}}%
\makeatletter\setlength{\@fptop}{0pt}\setlength{\@fpbot}{0pt plus 1fil}\makeatother
\begin{table}[htbp]
  \centering
  \caption{\textbf{Drama~3 control syntax.}}
  \label{table:syntax}
  \footnotesize
  \setlength{\tabcolsep}{2.5pt}
  \renewcommand{\arraystretch}{1.12}
  \begin{tabular}{@{}lll@{}}
    \toprule
    \textbf{Control} & \textbf{Syntax} & \textbf{Examples} \\
    \midrule
    \multicolumn{3}{@{}l}{\textit{Structure}} \\
    Speaker token &
      \begin{tabular}[t]{@{}l@{}}\texttt{<|speaker:N|>}, $N = 0,\dots,15$\end{tabular} &
      \begin{tabular}[t]{@{}l@{}}\texttt{<|speaker:0|> Ready? <|speaker:1|> Always.}\end{tabular} \\
    Reference block &
      \begin{tabular}[t]{@{}l@{}}\texttt{[Reference N]}\\ \texttt{Text: <|speaker:S|> transcript}\\ \texttt{Speech: \{\{audio\}\}}\end{tabular} &
      \begin{tabular}[t]{@{}l@{}}Figure~\ref{fig:prompt-clone}\end{tabular} \\
    Line separator &
      \begin{tabular}[t]{@{}l@{}}\texttt{<|audio\_separator|>}\end{tabular} &
      \begin{tabular}[t]{@{}l@{}}\texttt{The first line.}\\ \texttt{<|audio\_separator|>}\\ \texttt{The second line.}\end{tabular} \\
    Infill markers &
      \begin{tabular}[t]{@{}l@{}}\texttt{<|fim\_suffix|>\{\{after\}\}}\\ \texttt{<|fim\_prefix|>\{\{before\}\}}\end{tabular} &
      \begin{tabular}[t]{@{}l@{}}Figure~\ref{fig:prompt-infill}\end{tabular} \\
    Streaming look-ahead &
      \begin{tabular}[t]{@{}l@{}}\texttt{chunk<|endofturn|>next chunk}\end{tabular} &
      \begin{tabular}[t]{@{}l@{}}\texttt{It was late<|endofturn|>and the hall was empty.}\end{tabular} \\
    \midrule
    \multicolumn{3}{@{}l}{\textit{Direction}} \\
    Point tag &
      \begin{tabular}[t]{@{}l@{}}\texttt{[tag]} at a word position\end{tabular} &
      \begin{tabular}[t]{@{}l@{}}\texttt{Well [sighs] fine, you win. [laughs]}\\ \texttt{[pause] [gasps] [clears throat] [hesitates]}\end{tabular} \\
    Scoped tag &
      \begin{tabular}[t]{@{}l@{}}\texttt{<tag> ... </tag>}\end{tabular} &
      \begin{tabular}[t]{@{}l@{}}\texttt{Promise me <whisper>you won't tell her</whisper>.}\\ \texttt{<slow>...</slow>, <emphasis>...</emphasis>}\end{tabular} \\
    Style prefix &
      \begin{tabular}[t]{@{}l@{}}\texttt{[\ph{style}] line}\end{tabular} &
      \begin{tabular}[t]{@{}l@{}}\texttt{[with bright eager excitement] We won!}\\ \texttt{[icy calm, as if you already accepted it] Fine.}\end{tabular} \\
    Director's note &
      \begin{tabular}[t]{@{}l@{}}\texttt{Direction: \ph{note}}\\ \texttt{Line: \ph{text}}\end{tabular} &
      \begin{tabular}[t]{@{}l@{}}\texttt{Direction: icy calm, as if you already}\\ \texttt{\ \ accepted it. Put the rise on "goodbye".}\\ \texttt{Line: So this is goodbye.}\end{tabular} \\
    Relative note &
      \begin{tabular}[t]{@{}l@{}}\texttt{[\ph{change} versus previous] line}\\ (second and later turns)\end{tabular} &
      \begin{tabular}[t]{@{}l@{}}\texttt{[softer and warmer versus previous] ...}\\ \texttt{Direction: more amused than before;}\\ \texttt{\ \ drop the hurt.}\end{tabular} \\
    \midrule
    \multicolumn{3}{@{}l}{\textit{Hints}} \\
    Language hint &
      \begin{tabular}[t]{@{}l@{}}\texttt{line [language: xx]}\end{tabular} &
      \begin{tabular}[t]{@{}l@{}}\texttt{Welcome back to the show. [language: en]}\end{tabular} \\
    Duration hint &
      \begin{tabular}[t]{@{}l@{}}\texttt{line [audio tokens: N]}, $N$ in frames\end{tabular} &
      \begin{tabular}[t]{@{}l@{}}\texttt{Welcome back to the show. [audio tokens: 65]}\\ (about 3\,s)\end{tabular} \\
    \bottomrule
  \end{tabular}
\end{table}

\textbf{Conventions.}
A control is a short span in square or angle brackets, typically a few words, without nested brackets or line breaks; text in round parentheses is spoken.
The direction controls (point tags, scoped tags, style prefixes, director's notes, and relative notes) are free natural language, so any wording can be used and Table~\ref{table:syntax} only gives examples.
A point tag acts at its position, a scoped tag on the enclosed span, and a style prefix, director's note, or relative note on the whole line.
Language and duration hints condition the output and are never spoken; the language hint is optional and names the main language of a request, and each line is spoken in the language of its own text.

\end{document}